\documentclass[]{article}
\usepackage[english]{babel}
\usepackage{multind}
\usepackage{lineno}
\usepackage{graphicx,multicol}
\usepackage{epic,eepic,epsfig}
\usepackage{amssymb}
\begin{document}

\newcommand{\todo}[1]{\marginpar{{\textbf{JBJ}}\\\raggedright #1}}
\newcommand{\todoF}[1]{\marginpar{{\textbf{Fred}}\\\raggedright #1}}

\newcommand{\induce}[2]{\mbox{$ #1 \langle #2 \rangle$}}
\newcommand{\2}{\vspace{2mm}}
\newcommand{\dom}{\mbox{$\rightarrow$}}
\newcommand{\ndom}{\mbox{$\not\rightarrow$}}
\newcommand{\compdom}{\mbox{$\Rightarrow$}}
\newcommand{\cdom}{\compdom}
\newcommand{\sdom}{\mbox{$\Rightarrow$}}
\newcommand{\lsd}{locally semicomplete digraph}
\newcommand{\lt}{local tournament}
\newcommand{\qed}{\hfill$\diamond$}
\newcommand{\la}{\langle}
\newcommand{\ra}{\rangle}
\newcommand{\pf}{{\bf Proof: }}
\newtheorem{theorem}{Theorem}[section]
\newtheorem{problem}[theorem]{Problem}
\newtheorem{lemma}[theorem]{Lemma}
\newtheorem{proposition}[theorem]{Proposition}
\newtheorem{definition}[theorem]{Definition}
\newtheorem{corollary}[theorem]{Corollary}
\newtheorem{conjecture}[theorem]{Conjecture}
\newtheorem{claim}{Claim}
\newcommand{\beq}{\begin{equation}}
\newcommand{\eeq}{\end{equation}}
\newcommand{\<}[1]{\left\langle{#1}\right\rangle}

\newcommand{\Kij}[2]{K_{#1,#2}^{\rightarrow} }

\newcommand{\AY}[1]{{#1 }}
\newcommand{\JBJ}[1]{{#1}}

\bibliographystyle{plain}

\title{Completing orientations of partially oriented graphs}
\author{J. Bang-Jensen\thanks{Department of Mathematics and Computer Science, University of Southern Denmark, Odense DK-5230, Denmark (email:jbj@imada.sdu.dk).
 The research of Bang-Jensen was  supported by Canadian NSERC and by the Danish research council under grant number 1323-00178B. This work was carried out while the first author was visiting 
         Department of Mathematics and Statistics, 
         University of Victoria which is thanked for providing an excellent working environment.}
  \and J. Huang\thanks{Department of Mathematics and Statistics, University of Victoria,
P.O. BOX 3060 STN CSC, 
                  Victoria, B.C., V8W 3R4, Canada;
                  (email: huangj@uvic.ca). Research is supported by NSERC} \and 
X. Zhu\thanks{Department of Mathematics, Zhejiang Normal University, Jinhua,
     China; (email: xdzhu@zjnu.edu.cn).}}

\date{}
\maketitle

\begin{abstract}
We initiate a general study of what we call orientation completion problems.
For a fixed class $\cal C$ of oriented graphs, the orientation completion
problem asks whether a given partially oriented graph $P$ can be completed to
an oriented graph in $\cal C$ by orienting the (non-oriented) edges in $P$.
Orientation completion problems commonly generalize several existing
problems including recognition of certain classes of graphs and digraphs 
as well as extending representations of certain geometrically representable 
graphs.

We study orientation completion problems for various classes of oriented graphs,
including $k$-arc-strong oriented graphs, $k$-strong oriented graphs, 
quasi-transitive oriented graphs, local tournament, acyclic local tournaments,
locally transitive tournaments, locally transitive local tournaments, 
in-tournaments, and oriented graphs which have directed cycle factors.
We show that the orientation completion problem for each of these classes is
either polynomial time solvable or NP-complete. 
We also show that some of the NP-complete problems become polynomial time
solvable when the input oriented graphs satisfy certain extra conditions.
Our results imply that the representation extension problems for 
proper interval graphs and for proper circular arc graphs 
are polynomial time solvable, which generalize a previous result.

\noindent{}{\bf Keywords:} orientation completion problem, recognition,
representation extension, partially oriented graph, friendly partial oriented 
graph, local tournament, locally transitive local tournament, in-tournament, 
proper interval graph, proper circular arc graph, 
NP-complete, polynomial time algorithm.

\end{abstract}

\section{Introduction}

For a fixed class $\cal C$ of oriented graphs, the {\bf orientation completion 
problem} asks whether a given partially oriented graph can be completed to 
an oriented graph in $\cal C$ by orienting the (non-oriented) edges. 
Orientation completion problems commonly generalize several existing 
problems including recognition of certain classes of graphs and digraphs 
as well as extending representations of certain geometrically representable 
graphs. 

For the class of acyclic oriented graphs, the orientation completion problem 
is easily seen to be polynomial time solvable; a partially oriented graph can be
completed to an acyclic oriented graph if and only if it does not contain 
a directed cycle. 

When the fixed class $\cal C$ consists of all transitive oriented graphs, 
the underlying graphs of the oriented graphs in $\cal C$ are precisely 
the comparability graphs. Thus if the input partially oriented graph has 
no oriented edges the corresponding orientation completion problem just asks 
whether the input is a comparability graph.
So the problem is just the {\bf recognition problem} for comparability graphs,
cf. \cite{golumbicC18}. 
In general the orientation completion problem for this $\cal C$ can be solved 
in polynomial time using Gallai's decomposition scheme of comparability graphs,
cf. \cite{FKT,gallaiAMASH18}.

Suppose that $\cal C$ is the class of all strong oriented graphs.  
A result from \cite{boeschAMM87} implies that a partially oriented graph 
can be completed to a strong oriented graph if and only if it has no bridge and
no directed cut. Either a bridge or a directed cut in a partially oriented 
graph (if any exists) can be detected in polynomial time. Hence the orientation 
completion problem for $\cal C$ is polynomial time solvable. 
In contrast to this, the orientation completion problem for
the class of $k$-strong oriented graphs is NP-complete for each $k \geq 3$
(see Theorem \ref{k-conn}).

Orientation completion problems also generalize the problem of extending 
partial proper interval representation of proper interval graphs, 
cf. \cite{kkk}. A proper interval graph is the intersection graph of 
a family of intervals in a line where no interval is contained in another. 
Suppose that an induced subgraph $H$ of a graph $G$ is represented by a family
of intervals where no interval contains another. 
The {\bf representation extension problem} for proper interval graphs
asks whether it is possible to obtain a proper interval representation of $G$ 
that includes the one given for $H$. It is well-known that 
a proper interval representation of $G$ corresponds to 
an acyclic local tournament orientation $G$, cf. \cite{hellJGT20}. 
Thus the representation extension problem 
for proper interval graphs is just the orientation completion problem for 
the class of acyclic local tournaments where the partial orientation 
of $G$ corresponds to an interval representation of $H$.
The representation extension problem for proper interval graphs was shown to be 
polynomial time solvable, cf. \cite{kkk}.

In this paper we study orientation completion problems for various classes
of oriented graphs but focusing on three of them: local tournaments, 
locally transitive local tournaments, and acyclic local tournaments.
These three classes are nested; the class of local tournaments properly
contains the class of locally transitive local tournaments which in turn
properly contains the class of acyclic local tournaments. 
We show that the complexity of the orientation problems for 
the three classes alternates; while the orientation completion problem 
is polynomial time solvable for the class of local tournaments and 
for the class of acyclic local tournaments, it is NP-complete for the class
of locally transitive local tournaments. In fact we show that the problem 
remains NP-complete for the class of locally transitive tournaments
(i.e., complete locally transitive local tournaments). 
Since, as mentioned above, the orientation completion problem for 
acyclic local tournaments generalizes the representation extension problem 
for proper interval graphs, our result on the orientation completion problem 
for acyclic local tournaments generalizes a result from \cite{kkk}.
We also show that if the input oriented graphs are restricted to 
being friendly (see the definition in Section~\ref{polys}) then 
the orientation completion problem for locally transitive local tournaments 
is polynomial time solvable. 
The underlying graphs of connected locally transitive local tournaments are 
precisely the proper circular arc graphs.
Our result on the restricted orientation completion problem for 
locally transitive local tournaments implies that 
the representation problem for proper circular arc graphs is solvable in
polynomial time.
For various other classes of oriented graphs, including 
$k$-arc-strong oriented graphs, $k$-strong oriented graphs, quasi-transitive
oriented graphs, in-tournaments, 
and oriented graphs which have directed cycle factors, we show that
the corresponding orientation completion problems are either polynomial 
time solvable or NP-complete. 

The paper is organized as follows. In Section~\ref{terms} we provide some 
terminology and notation as well as some preliminary results. 
We show in Section~\ref{npc} that two polynomially equivalent problems are
NP-complete. These include the orientation completion problem for 
the class of locally transitive tournaments. 
Section~\ref{polys} includes several cases where the orientation problems
are shown to be polynomial time solvable. 
Finally in Section~\ref{remarks} we make some concluding remarks and 
state some open problems.

\section{Terminology and preliminary results}
\label{terms}

Notation or terminology not introduced here follows \cite{bang2009}. 
Both graphs and digraphs will be considered in this paper. For graphs we 
assume that they do not contain loops or multiple edges (i.e., they are 
{\bf simple}) and for digraphs we assume they do not contain loops or 
two arcs joining the same pair of vertices (i.e., they are {\bf oriented graphs}).
Most of the time we will consider the so-called partially oriented graphs
which may contain both edges and arcs.

A {\bf partially oriented graph (pog)} $P=(V,E\cup A)$ is the edge-disjoint
union of a graph $G=(V,E)$ and an oriented graph $D=(V,A)$ on the same vertex
set $V$; thus no two vertices in a partially oriented graph can have two
vertices joined by both an edge and an arc. In the case when 
$A = \emptyset$, $P$ is just the graph $G$ and similarly when 
$E = \emptyset$, $P$ is the oriented graph $D$.
When there is either an edge or an arc joining two vertices $u, v$ in $P$, 
we say that $u, v$ are {\bf adjacent}, and use $uv$ to denote the edge between 
$u$ and $v$ and use $(u,v)$ to denote the arc from $u$ to $v$.
If $(x,y)$ is an arc of $P$ we say that $x$ {\bf dominates} $y$ and call $y$ an 
{\bf out-neighbour} of $x$ and $x$ an {\bf in-neighbour} of $y$.  
For a vertex $v\in V(D)$ we denote by $N^+(v)$ ($N^-(v)$) the set of 
out-neighbours (in-neighbours) of $v$.
The {\bf underlying graph} of a pog $P$, denoted $UG(P)$,
is obtained from $P$ by suppressing the orientation
of each arc.  A pog $P$ is {\bf connected} if $UG(P)$ is a connected graph.

Given a partially oriented graph $P = (V,E\cup A)$, {\bf to complete} $P$
means to obtain an oriented graph by orienting each edge in $E$ (that is,
replacing each edge by an arc in one of the two possible ways). 
The central topic of this paper is the study of the following problem:
Let $\cal C$ be a fixed class of oriented graphs.

\medskip

\fbox{\begin{minipage}{20em}

{\bf Orientation Completion Problem for $\cal C$.} 

\begin{tabbing}
\hspace{0.5cm} \= {\bf Instance:}\ \ \ \= A partially oriented graph $P$.\\
\>  {\bf Question:} \> Can $P$ be completed to an oriented graph in $\cal C$?
\end{tabbing}

\end{minipage}}

\medskip

As an example suppose that $\cal C$ consists of all transitive tournaments
(i.e., tournaments that contain no directed triangle).  
Then the orientation problem for $\cal C$ asks whether a partially oriented 
graph $P$ can be completed to a transitive tournament.
Clearly, if $UG(P)$ contains a pair of non-adjacent vertices or $P$ contains 
a directed cycle, then $P$ cannot be completed to a transitive tournament.
A pair of non-adjacent vertices in $UG(P)$ or a directed cycle in $P$ 
(if one exists) can be found in polynomial time. Otherwise $P$ can be completed 
to a tournament in $\cal C$ as shown below.

\begin{proposition}
\label{acyclicext}
Let $P = (V, A \cup E)$ be a partially oriented graph. If $UG(P)$ does not 
contain a pair of non-adjacent vertices and $P$ does not contain 
a directed cycle, then $P$ can be completed to a transitive tournament
\end{proposition}

\pf Since $P$ does not contain a directed cycle, its vertices can be ordered 
$v_1,v_2,\ldots{},v_n$ so that each arc $(v_i,v_j) \in A$ satisfies that $i<j$. 
Now we can now replace each edge $v_kv_{\ell} \in E$ with $k < \ell$ by
the arc $(v_k,v_{\ell})$. Since $UG(P)$ does not contain a pair of non-adjacent
vertices, it is complete and the resulting oriented graph is a transitive 
tournament completion of $P$.
\qed

\medskip

A {\bf local tournament} is an oriented graph $D$ such that for every vertex $v$,
$N^-(v)$ and $N^+(v)$ each induces a tournament in $D$. If $N^-(v)$ and $N^+(v)$ 
each induces a transitive tournament for every vertex $v$, then $D$ is called
{\bf locally transitive}. It follows from these definitions that every tournament
is a local tournament and every transitive tournament is 
a locally transitive local tournament 
\footnote{Locally transitive local tournaments are previously 
called {\bf local transitive tournaments}, cf. \cite{huangJCT63}}.
But neither of the converses is true, as local tournaments or
locally transitive local tournaments may not be tournaments. 
When a local tournament is also a tournament, we call it {\bf complete}. 
When a locally transitive local tournament is complete, we call it
a {\bf locally transitive tournament}.
 
A graph $G = (V,E)$ is a {\bf proper circular arc graph} if there is a family 
of circular arcs $J_v, v \in V$ on a circle such that no circular arc is 
contained in another and for any two vertices $u, v \in V$, $uv \in E$ 
if and only if $J_u \cap J_v \neq \emptyset$. Such a family of circular arcs is 
called a {\bf proper circular arc representation} of $G$.  
Given a proper circular arc representation $J_v, v \in V$ of $G = (V,E)$,
an orientation of $G$ can be obtained in such a way that $(u,v)$ is an
arc if and only if $J_u$ contains the counterclockwise endpoint of $J_v$.
It is easy to see that this orientation of $G$ is locally transitive and
hence is a local tournament. Thus every proper circular arc graph can 
be completed to a local tournament. 
The following theorem due to Skrien \cite{skrienJGT6} assures that 
the converse is also true for connected graphs.

\begin{theorem} \cite{skrienJGT6} \label{relation}
Let $G$ be a connected graph. The following statements are equivalent:
\begin{enumerate}
\item $G$ can be completed to a local tournament;
\item $G$ can be completed to a locally transitive local tournament;
\item $G$ is a proper circular arc graph.
\qed
\end{enumerate}
\end{theorem}    

Proper interval graphs form a subclass of proper circular arc graphs.
A graph $G = (V,E)$ is a {\bf proper interval graph} if there is a family of 
intervals $I_v, v \in V$ on a line such that no interval is contained in another
and for any two vertices $u, v \in V$, $uv \in E$ if and only if
$I_u \cap I_v \neq \emptyset$. Such a family of intervals is called 
a {\bf proper interval representation} of $G$. Given a proper interval 
representation of $G$, one can obtain an acyclic local tournament orientation of 
$G$ is a similar way as above for proper circular arc graphs. 
Conversely, if $G$ is oriented as an acyclic local tournament then 
a proper interval graph representation of $G$ can be obtained from
the orientation, cf. \cite{hellJGT20}.

\begin{theorem} \cite{hellJGT20} \label{ship}
Let $G$ be a graph. The following statements are equivalent:
\begin{enumerate}
\item $G$ can be completed to an acyclic local tournament;
\item $G$ is a proper interval graph.
\qed
\end{enumerate}
\end{theorem}

A {\bf round ordering} of a digraph $D$ is a cyclic ordering 
${\cal O}=v_1,v_2,\ldots{},v_n,v_1$ of the vertices of $D$ 
such that for each vertex $v_i$ we have 
$N^+(v_i)=\{v_{i+1},\ldots{},v_{d^+(v_i)+i}\}$ and 
$N^-(v_i)=\{v_{i-d^-(v_i)},\ldots{},v_{i-1}\}$ where indices are modulo $n$. 
A digraph which has a round ordering is called {\bf round}.
Round digraphs are characterized in \cite{huangAJC19}.
It is easy to see that if an oriented graph has a round ordering then
it is locally transitive. The following theorem asserts that the converse
is also true.

\begin{theorem}\cite{bangJGT14} \label{roundLT}
An oriented graph $D$ has a  round  ordering 
${\cal O}=v_1,v_2,\ldots{},v_n,v_1$ of its vertices if and only if $D$ is 
a local tournament which is locally transitive. Furthermore, there is 
a polynomial algorithm for deciding whether a given oriented graph is round 
and find a round ordering if one exists. 
\qed
\end{theorem}

Let $D$ be an oriented graph on the vertices $v_1, v_2, \dots, v_n$ and 
let $D_1, D_2, \dots, D_n$ be vertex disjoint oriented graphs. 
{\bf To substitute $D_i$ for $v_i$} for each $i \in [n]$ is to obtain 
a new oriented graph from $D$ by replacing $v_i$ with $D_i$ such that 
if $(v_i,v_j)$ is an arc in $D$ then $(x,y)$ is an arc in $D'$ for all
$x \in V(D_i)$ and $y \in V(D_j)$ and there is no other arcs in $D'$.
We use $D[D_1,D_2,\dots,D_n]$ to denote the new digraph $D'$.
When $D$ is locally transitive and $D_i$ is a transitive tournament 
for each $i \in [n]$, $D'$ is also locally transitive.

A tournament $T$ is {\bf highly regular} if it is regular and has a round 
ordering. Highly regular tournaments are the basic structure frames for 
constructing all tournaments that are locally transitive. 

\begin{theorem} \cite{moonCMB21} \label{moon}
Every locally transitive tournament is obtained from a highly regular
tournament $T$ by substituting a transitive tournament for each vertex of $T$.
\qed
\end{theorem}

Given a locally transitive tournament $T'$, we can obtain a highly regular
tournament $T$ as indicated in Theorem \ref{moon} by repeatedly identifying
two vertices $x, y$ with $N^-(x) \setminus \{x,y\} = N^-(y) \setminus \{x,y\}$
and $N^+(x) \setminus \{x,y\} = N^+(y) \setminus \{x,y\}$ and deleting 
the loop resulted from the identification, until no two such vertices remain.

Let ${\cal O}=v_1,v_2,\ldots{},v_n,v_1$ be a cyclic ordering of the vertices of a pog $P=(V,E\cup A)$. An arc $(v_i,v_j)\in A$ {\bf dominates} an arc $(v_s,v_t)\in A$ with respect to $\cal O$ if the vertices of the two arcs appear in the order $v_i,v_s,v_t,v_j$ in ${\cal O}$, where we can have $i=s$ or $j=t$. An arc $(v_i,v_j)\in A$ dominates an edge $v_pv_q$ if both of the vertices $v_p,v_q$ occur in the interval $[v_i,v_j]$ from $v_i$ to $v_j$ according to $\cal O$. An arc is {\bf maximal} with respect to $\cal O$ if it is not dominated by any other arc. A cyclic ordering of the vertices of a pog $P$  is {\bf excellent} if $P$ has no pair of arcs $(v_i,v_j), (v_s,v_t)$ so that these vertices occur in the order $v_i,v_t,v_s,v_j$ in the cyclic ordering, where we may have $i=t$ or $s=j$. 

\begin{lemma}\label{completeexcellent}
Suppose $P=(V,E\cup A)$  is pog which  has  an excellent cyclic ordering  ${\cal O}=v_1,\ldots{},v_n,v_1$ of its vertices.
Then $P$ can be completed to an oriented graph $D'$ for which  the same  cyclic ordering $\cal O$ is excellent.
\end{lemma}

\pf Let $P=(V,E\cup A)$  be a  pog and let ${\cal O}=v_1,v_2,\dots{},v_n,v_1$ be an excellent cyclic ordering of $D$. Let 
$a_1=(v_{i_1},v_{j_1}),a_2=(v_{i_2},v_{j_2}),\ldots{},a_k=(v_{i_k},v_{j_k})$ be the maximal arcs of $D$ with respect to $\cal O$. By the assumption of the lemma, for each arc $a_r$ every  arc $(v_p,v_q)$ for which both  vertices $v_p,v_q$ occur after 
in the interval $[v_i,v_j]$  satisfy that the vertices occur in the order $v_{i_r},v_p,v_q,v_{j_r}$.
For each $r\in [k]$ in increasing order and all indices $p,q$ with 
$v_{i_r},v_p,v_q,v_{j_r}$ occurring in that order such that $v_pv_q$ is an edge of $P$ we orient this edge as the arc $(v_p,v_q)$. Let $D^*=(V,A\cup A^*)$ be the oriented graph consisting of the original arcs and those edges which we have oriented so far. By construction of $D^*$, $\cal O$ is an excellent ordering of $D^*$.
Hence if no edge of $E$ is still unoriented we are done. It suffices to show that we may orient one of the remaining edges, since then the claim follows by induction on the number of unoriented edges. Let $v_pv_q$ be an edge which was not oriented and orient this as $(v_p,v_q)$. We claim that $\cal  O$ is an excellent ordering of $D^*\cup \{(v_p,v_q)\}$. If not then there is an arc $(v_a,v_b)$ of $D^*$  such that the vertices occur in the order $v_p,v_b,v_a,v_q$ but then the edge $v_pv_q$ is dominated by the arc $(v_a,v_b)$ and hence by one of the arcs $a_1,\ldots{},a_k$, contradicting that it was not oriented above. \qed

\begin{lemma}
\label{excellentextend}
 An oriented graph $D$  has an excellent cyclic ordering $\cal O$ if and only if it can be extended to a round local tournament $D^*$ by adding new arcs. In particular, every excellent ordering of $D$ is a round ordering of $D^*$ and conversely.
\end{lemma}

\pf Suppose first that $D$ can be extended to a round  local tournament $D^*$. According to Theorem \ref{roundLT} there is a round  ordering ${\cal O}=v_1,v_2,\ldots{},v_n,v_1$ of $V(D^*)=V(D)$.  We claim that this ordering is also excellent. If not, then there are arcs $(v_i,v_j)$ and $(v_s,v_t)$ so that the vertices occur in the order $v_i,v_t,v_s,v_j$ according to $\cal O$. Since $\cal O$ is a round ordering, we have that $(v_i,v_t)$ and $(v_t,v_j)$ are arcs of $D^*$ but then the neighbours of $v_t$ do not occur correctly according to $\cal O$, contradiction. So $\cal O$ is an excellent ordering of $D^*$ and hence also of the subdigraph $D$.
To prove the second part let ${\cal O}=v_1,v_2,\dots{},v_n,v_1$ be an excellent cyclic ordering of the oriented graph $D$. It suffices to observe that for every maximal arc $(v_i,v_j)$ with respect to $\cal O$ and any pair of non-adjacent vertices
$v_a,v_b$ in the interval $[v_i,v_j]$ with $v_a$ before $v_b$ we may add the arc $(v_a,v_b)$ and still have an excellent ordering of the resulting oriented graph. Now the claim follows by induction on the number of such non-adjacent pairs. 
\qed

\begin{lemma}\label{roundtoT}
Every round oriented graph $D$ can be completed to 
a locally transitive tournament.
\end{lemma}

\pf We prove the statement by induction on the number of vertices in $D$.
which are not adjacent to all other vertices.
By Theorem \ref{roundLT}, the base case where there is  no such vertex is true. So assume that all round oriented graph on $n$ vertices with at most $k$ vertices as above can be completed to a locally transitive tournament and let $D$ be a round digraph with $k+1$ vertices each of which has  a non-neighbour. Let ${\cal O}=v_1,v_2,\dots{},v_n,v_1$ be a round  ordering according to Theorem \ref{roundLT}. W.l.o.g. the vertex $v_1$ has a non-neighbour, so we have that $v_{d^+(v_1)+2}\neq v_{n-d^-(v_1)}$. We claim that there is no arc $(v_p,v_q)$ with $1\leq q<p<n-d^-(v_1)$. Suppose such an arc does exist. Then we have $p>d^+(v_1)+1$ by the choice of $\cal O$ and we have $q>1$ since $v_p$ is not adjacent to $v_1$. But this contradicts the fact that the vertex $v_p$ sees its out-neighbourhood as an interval just after itself according to $\cal O$ because $v_1$ is not-adjacent to $v_p$. Thus if add all the arcs $(v_1,v_{d^+(v_1)+2}),\ldots{}, (v_1,v_{n-d^-(v_1)-1})$ to $D$ the order $\cal O$ is an excellent ordering of the resulting digraph $D'$. By Lemmas \ref{completeexcellent} and \ref{excellentextend} this implies that $D'$ can be extended to a round local 
tournament $D''$ by adding new arcs. Now the claim follows by induction since $D'$ has less vertices with non-neighbours that $D$ does. \qed


\section{Hardness results on excellent orderings and completions to locally transitive tournaments}
\label{npc}

For a given oriented graph $D$ we denote by $D^c$ the partially oriented complete
graph obtained from $D$ by adding an edge between each pair of non-adjacent 
vertices.
We shall use the following consequence of Lemmas \ref{completeexcellent}, \ref{excellentextend} and \ref{roundtoT} as well as Theorem \ref{roundLT}.

\begin{lemma}
\label{excellentLTT}
An oriented graph $D$ has an excellent ordering if and only if the pog $D^c$ has a completion $T$ to a tournament which is locally transitive. Furthermore, given an excellent ordering of $D$ we can construct $T$ in polynomial time and conversely, given $T$ we can obtain an excellent ordering of $D$.\qed
\end{lemma}


\begin{figure}[h!]
\begin{center}
\setlength{\unitlength}{0.00034996in}
\begingroup\makeatletter\ifx\SetFigFont\undefined%
\gdef\SetFigFont#1#2#3#4#5{%
  \reset@font\fontsize{#1}{#2pt}%
  \fontfamily{#3}\fontseries{#4}\fontshape{#5}%
  \selectfont}%
\fi\endgroup%
{\renewcommand{\dashlinestretch}{30}
\begin{picture}(8636,3511)(0,-10)
\put(105,3034){\blacken\ellipse{180}{180}}
\put(105,3034){\ellipse{180}{180}}
\put(3030,3034){\blacken\ellipse{180}{180}}
\put(3030,3034){\ellipse{180}{180}}
\put(105,784){\blacken\ellipse{180}{180}}
\put(105,784){\ellipse{180}{180}}
\put(3030,784){\blacken\ellipse{180}{180}}
\put(3030,784){\ellipse{180}{180}}
\put(5055,3034){\blacken\ellipse{180}{180}}
\put(5055,3034){\ellipse{180}{180}}
\put(5055,784){\blacken\ellipse{180}{180}}
\put(5055,784){\ellipse{180}{180}}
\put(7980,3034){\blacken\ellipse{180}{180}}
\put(7980,3034){\ellipse{180}{180}}
\put(7980,784){\blacken\ellipse{180}{180}}
\put(7980,784){\ellipse{180}{180}}
\path(105,3034)(2895,3034)
\path(105,3034)(2895,3034)
\blacken\path(2775.000,3004.000)(2895.000,3034.000)(2775.000,3064.000)(2775.000,3004.000)
\path(150,3034)(2940,874)
\path(150,3034)(2940,874)
\blacken\path(2826.748,923.739)(2940.000,874.000)(2863.478,971.182)(2826.748,923.739)
\path(3030,3034)(195,874)
\path(3030,3034)(195,874)
\blacken\path(272.271,970.588)(195.000,874.000)(308.633,922.862)(272.271,970.588)
\path(105,784)(2895,784)
\path(105,784)(2895,784)
\blacken\path(2775.000,754.000)(2895.000,784.000)(2775.000,814.000)(2775.000,754.000)
\path(105,3034)(105,784)
\path(105,3034)(105,784)
\path(3030,3034)(3030,739)
\path(3030,3034)(3030,739)
\path(5055,2989)(5055,784)
\path(7980,3034)(7980,829)
\path(5100,829)(7890,2989)
\blacken\path(7813.478,2891.818)(7890.000,2989.000)(7776.748,2939.261)(7813.478,2891.818)
\path(5010,784)(7800,784)
\blacken\path(7680.000,754.000)(7800.000,784.000)(7680.000,814.000)(7680.000,754.000)
\path(7980,3034)(5145,3034)
\blacken\path(5265.000,3064.000)(5145.000,3034.000)(5265.000,3004.000)(5265.000,3064.000)
\path(5055,3034)(7935,829)
\blacken\path(7821.482,878.129)(7935.000,829.000)(7857.957,925.769)(7821.482,878.129)
\put(15,424){\makebox(0,0)[lb]{\smash{{$b$}}}}
\put(60,3304){\makebox(0,0)[lb]{\smash{{$a$}}}}
\put(2940,3349){\makebox(0,0)[lb]{\smash{{$\alpha$}}}}
\put(3210,469){\makebox(0,0)[lb]{\smash{{$\beta$}}}}
\put(7890,3349){\makebox(0,0)[lb]{\smash{{$\alpha$}}}}
\put(8160,469){\makebox(0,0)[lb]{\smash{{$\beta$}}}}
\put(4965,3304){\makebox(0,0)[lb]{\smash{{$u$}}}}
\put(5010,379){\makebox(0,0)[lb]{\smash{{$v$}}}}
\put(1410,19){\makebox(0,0)[lb]{\smash{{$X$}}}}
\put(6405,64){\makebox(0,0)[lb]{\smash{{$\bar{X}$}}}}
\end{picture}
}

\end{center}
\caption{Two different labellings of the same  partially oriented complete graph on 4 vertices. For later convenience we name these  $X,\bar{X}$.}\label{varfixfig}
\end{figure}
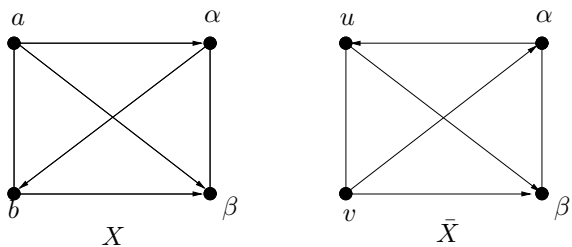

The following is easy to check.

\begin{proposition}\label{fixvar}
Each of the two labellings $X,\bar{X}$ partially oriented complete graphs $X,\bar{X}$ in Figure \ref{varfixfig} have exactly two completions to a locally transitive tournament. For $X$ these are obtained by orienting the two edges $ab,\alpha{}\beta$ as either $(b,a),(\beta,\alpha)$ or $(a,b),(\alpha,\beta)$. For $\bar{X}$ they are obtained  by orienting the two edges $uv,\alpha{}\beta$ as either $(v,u),(\alpha,\beta)$ or 
$(u,v),(\beta,\alpha)$. \qed
\end{proposition}

\begin{figure}[h!]
\begin{center}
\setlength{\unitlength}{0.00034996in}
\begingroup\makeatletter\ifx\SetFigFont\undefined%
\gdef\SetFigFont#1#2#3#4#5{%
  \reset@font\fontsize{#1}{#2pt}%
  \fontfamily{#3}\fontseries{#4}\fontshape{#5}%
  \selectfont}%
\fi\endgroup%
{\renewcommand{\dashlinestretch}{30}
\begin{picture}(5256,3600)(0,-10)
\put(1680,3120){\blacken\ellipse{180}{180}}
\put(1680,3120){\ellipse{180}{180}}
\put(3705,3120){\blacken\ellipse{180}{180}}
\put(3705,3120){\ellipse{180}{180}}
\put(4830,1770){\blacken\ellipse{180}{180}}
\put(4830,1770){\ellipse{180}{180}}
\put(3705,420){\blacken\ellipse{180}{180}}
\put(3705,420){\ellipse{180}{180}}
\put(1680,420){\blacken\ellipse{180}{180}}
\put(1680,420){\ellipse{180}{180}}
\put(555,1770){\blacken\ellipse{180}{180}}
\put(555,1770){\ellipse{180}{180}}
\put(2670,1770){\blacken\ellipse{180}{180}}
\put(2670,1770){\ellipse{180}{180}}
\path(1680,3120)(3705,3120)
\path(1680,3120)(3705,3120)
\path(4830,1770)(3705,420)
\path(4830,1770)(3705,420)
\path(1680,375)(555,1770)
\path(1680,375)(555,1770)
\path(3705,3120)(4785,1860)
\path(3705,3120)(4785,1860)
\blacken\path(4684.127,1931.587)(4785.000,1860.000)(4729.683,1970.635)(4684.127,1931.587)
\path(3750,420)(1770,420)
\path(3750,420)(1770,420)
\blacken\path(1890.000,450.000)(1770.000,420.000)(1890.000,390.000)(1890.000,450.000)
\path(555,1725)(1635,3030)
\path(555,1725)(1635,3030)
\blacken\path(1581.604,2918.426)(1635.000,3030.000)(1535.380,2956.680)(1581.604,2918.426)
\path(2670,1770)(1725,3030)
\path(2670,1770)(1725,3030)
\blacken\path(1821.000,2952.000)(1725.000,3030.000)(1773.000,2916.000)(1821.000,2952.000)
\path(2670,1770)(3660,3030)
\path(2670,1770)(3660,3030)
\blacken\path(3609.451,2917.107)(3660.000,3030.000)(3562.272,2954.176)(3609.451,2917.107)
\path(2670,1770)(4695,1770)
\path(2670,1770)(4695,1770)
\blacken\path(4575.000,1740.000)(4695.000,1770.000)(4575.000,1800.000)(4575.000,1740.000)
\path(2670,1770)(3615,510)
\path(2670,1770)(3615,510)
\blacken\path(3519.000,588.000)(3615.000,510.000)(3567.000,624.000)(3519.000,588.000)
\path(2670,1770)(1770,510)
\path(2670,1770)(1770,510)
\blacken\path(1815.337,625.085)(1770.000,510.000)(1864.161,590.211)(1815.337,625.085)
\path(2670,1770)(690,1770)
\path(2670,1770)(690,1770)
\blacken\path(810.000,1800.000)(690.000,1770.000)(810.000,1740.000)(810.000,1800.000)
\put(2985,1860){\makebox(0,0)[lb]{\smash{{$c$}}}}
\put(1500,3435){\makebox(0,0)[lb]{\smash{{$c_{11}$}}}}
\put(3615,3435){\makebox(0,0)[lb]{\smash{{$c_{12}$}}}}
\put(4965,1725){\makebox(0,0)[lb]{\smash{{$c_{21}$}}}}
\put(3705,60){\makebox(0,0)[lb]{\smash{{$c_{22}$}}}}
\put(1500,15){\makebox(0,0)[lb]{\smash{{$c_{31}$}}}}
\put(15,1680){\makebox(0,0)[lb]{\smash{{$c_{32}$}}}}
\end{picture}
}

\end{center}
\caption{A partially oriented wheel $W$}\label{clausefig}
\end{figure}
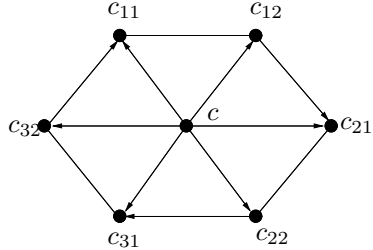

\begin{lemma}
\label{atleast one back}
Let $W$ be the partially oriented 6-wheel in Figure \ref{clausefig}. If we orient the three edges $c_{11}c_{12},c_{21}c_{22},c_{31}c_{32}$ as $(c_{11},c_{12}),(c_{21},c_{22}),(c_{31},,c_{32})$ the resulting digraph $D$ is not a  spanning subdigraph of any locally transitive tournament on 7 vertices and hence $D$ has no excellent ordering . For all the 7 remaining orientations of the three edges $c_{11}c_{12},c_{21}c_{22},c_{31}c_{32}$ the resulting digraph $D'$ can be completed to a locally transitive tournament and hence has an excellent ordering.
\end{lemma}

\pf If we orient the three edges as $(c_{11},c_{12}),(c_{21},c_{22}),(c_{31},c_{32})$ then the vertex $c$ has a directed 6-cycle in its out-neighbourhood and hence no completion to a locally transitive tournament exists. By Lemma \ref{excellentextend} this implies that $D$ has no excellent ordering. On the other hand, for each of the remaining 7 orientations of the three arcs, the digraph $D'$ we obtain is acyclic and hence it follows from Proposition \ref{acyclicext} that we can extend $D'$ to a transitive tournament $T$. The ordering corresponding to decreasing 
out-degrees of $T$ gives an excellent ordering of $W$.  \qed

\begin{theorem}
\label{LTTexthard}
The following polynomially equivalent problems are NP-complete.
\begin{itemize}
\item Deciding whether an oriented graph has an excellent ordering.
\item Deciding whether a given partially oriented complete graph can be 
      completed to a locally transitive tournament.
\end{itemize}
\end{theorem}

\pf We describe polynomial reductions from 3-SAT to these problems.

Let $\cal F$ be an instance of 3-SAT with variables $x_1,x_2,\ldots{},x_n$ and clauses $C_1,C_2,\ldots{},C_m$, where each clause is of the form $(\ell_1\vee\ell_2\vee\ell_3)$ and each $\ell_i$ is either one of the variables $x_j$ or the negation $\bar{x}_j$ of such a variable.

Let $p_i$ ($q_i$) be the number of times variable $x_i$ ($\bar{x}_i$) 
occurs as a literal in $\cal F$. 
The enumeration of the clauses $C_1,\ldots{},C_m$ induces an ordering on 
the occurrences of the same literal in the formula.
Guided by this ordering we  now construct a partially oriented graph $H'=H'({\cal F})$ as follows:

Let $X,\bar{X}$ be as in Figure \ref{varfixfig}.
For each variable $x_i$ we form the partially oriented graph $X_i$ from $p_i$ copies of $X$ and $q_i$  copies of $\bar{X}$ (these $p_i+q_i$ graphs are vertex disjoint) by identifying all the $\alpha$ vertices and all the $\beta$ vertices and denote  these identified vertices by $\alpha{}(x_i),\beta{}(x_i)$, respectively. Denote the $p_i$ copies of $a,b$ by $a_{i,1},\ldots{},a_{i,p_i},b_{i,1},
\ldots{},b_{i,p_i}$ and the $q_i$ copies of $u,v$ by $u_{i,1},\ldots{},u_{i,q_i}, v_{i,1},\ldots{},v_{i,q_i}$.

Take $m$ disjoint copies $W_1,W_2,\ldots{},W_m$ of the partially oriented 6-wheel from Figure \ref{clausefig} where the vertices of $W_i$ are denoted $c_i,c^i_{11},c^i_{12},c^i_{21},c^i_{22},c^i_{31},c^i_{32}$.
Make the following association between literals of $\cal F$ and the $W_i$'s: If $C_i=(\ell_{i,1}\vee\ell_{i,2}\vee\ell_{i,3})$ we associate the vertices $c^i_{j1},c^i_{j2}$ with the literal $\ell_{i,j}$ of $C_i$, $j\in [3]$.

Now we make the following vertex identifications. For each clause $C_i=(\ell_{i,1}\vee\ell_{i,2}\vee\ell_{i,3})$ we identify the vertices 
$c^i_{11},c^i_{12},c^i_{21},c^i_{22},c^i_{31},c^i_{32}$ with vertices from the union of the graphs $X_1,\ldots{},X_n$ as follows: If $\ell_{i,j}=x_r$ and this is the $h$'th occurrence of variable $x_r$ according to the induced ordering of that literal, then identify $c^i_{j1}$ with $a_{r,h}$ and $c^i_{j2}$ with $b_{r,h}$. If $\ell_{i,j}=\bar{x}_r$ and this is the $t$'th occurrence of $\bar{x}_r$ according to the induced ordering of that literal, then identify $c^i_{j1}$ with $u_{r,t}$ and $c^i_{j2}$ with $v_{r,t}$. Note that even after these identifications each of the subdigraphs $W_1,\ldots{},W_m$ are still vertex disjoint.

Clearly we can construct $H'$ in polynomial time from $\cal F$.
Denote by $H$ the oriented graph obtained from $H'$ by deleting all 
(unoriented) edges.
It is easy to check that the in- and out-neighbourhoods of each vertex in  $H$ is acyclic.

\begin{figure}[h!]
\begin{center}
\setlength{\unitlength}{0.0006in}
\begingroup\makeatletter\ifx\SetFigFont\undefined%
\gdef\SetFigFont#1#2#3#4#5{%
  \reset@font\fontsize{#1}{#2pt}%
  \fontfamily{#3}\fontseries{#4}\fontshape{#5}%
  \selectfont}%
\fi\endgroup%
{\renewcommand{\dashlinestretch}{30}
\begin{picture}(7061,5986)(0,-10)
\put(825,3124){\blacken\ellipse{128}{128}}
\put(825,3124){\ellipse{128}{128}}
\put(1275,3574){\blacken\ellipse{128}{128}}
\put(1275,3574){\ellipse{128}{128}}
\put(1725,4024){\blacken\ellipse{128}{128}}
\put(1725,4024){\ellipse{128}{128}}
\put(2175,4474){\blacken\ellipse{128}{128}}
\put(2175,4474){\ellipse{128}{128}}
\put(2625,4924){\blacken\ellipse{128}{128}}
\put(2625,4924){\ellipse{128}{128}}
\put(3075,5374){\blacken\ellipse{128}{128}}
\put(3075,5374){\ellipse{128}{128}}
\put(600,4924){\blacken\ellipse{128}{128}}
\put(600,4924){\ellipse{128}{128}}
\put(1050,5374){\blacken\ellipse{128}{128}}
\put(1050,5374){\ellipse{128}{128}}
\put(3975,5374){\blacken\ellipse{128}{128}}
\put(3975,5374){\ellipse{128}{128}}
\put(4425,4924){\blacken\ellipse{128}{128}}
\put(4425,4924){\ellipse{128}{128}}
\put(4875,4474){\blacken\ellipse{128}{128}}
\put(4875,4474){\ellipse{128}{128}}
\put(5325,4024){\blacken\ellipse{128}{128}}
\put(5325,4024){\ellipse{128}{128}}
\put(5775,3574){\blacken\ellipse{128}{128}}
\put(5775,3574){\ellipse{128}{128}}
\put(6225,3124){\blacken\ellipse{128}{128}}
\put(6225,3124){\ellipse{128}{128}}
\put(1275,1774){\blacken\ellipse{128}{128}}
\put(1275,1774){\ellipse{128}{128}}
\put(6000,5599){\blacken\ellipse{128}{128}}
\put(6000,5599){\ellipse{128}{128}}
\put(6000,5599){\blacken\ellipse{128}{128}}
\put(6000,5599){\ellipse{128}{128}}
\put(6450,5149){\blacken\ellipse{128}{128}}
\put(6450,5149){\ellipse{128}{128}}
\put(2130,1774){\blacken\ellipse{128}{128}}
\put(2130,1774){\ellipse{128}{128}}
\put(5550,1774){\blacken\ellipse{128}{128}}
\put(5550,1774){\ellipse{128}{128}}
\put(4875,1774){\blacken\ellipse{128}{128}}
\put(4875,1774){\ellipse{128}{128}}
\put(3975,1774){\blacken\ellipse{128}{128}}
\put(3975,1774){\ellipse{128}{128}}
\put(3075,1774){\blacken\ellipse{128}{128}}
\put(3075,1774){\ellipse{128}{128}}
\put(3975,649){\blacken\ellipse{128}{128}}
\put(3975,649){\ellipse{128}{128}}
\put(3075,649){\blacken\ellipse{128}{128}}
\put(3075,649){\ellipse{128}{128}}
\path(600,4924)(1050,5374)
\path(600,4924)(1050,5374)
\path(6000,5599)(6495,5104)
\path(6000,5599)(6495,5104)
\path(3075,649)(3975,649)
\path(3075,649)(3975,649)
\path(825,3079)(1275,3574)
\path(825,3079)(1275,3574)
\path(2625,4924)(3075,5374)
\path(2625,4924)(3075,5374)
\path(3975,5374)(4425,4879)
\path(3975,5374)(4425,4879)
\path(4875,4474)(5325,3979)
\path(4875,4474)(5325,3979)
\path(5775,3574)(6270,3079)
\path(5775,3574)(6270,3079)
\path(1275,1774)(2130,1774)
\path(1275,1774)(2130,1774)
\path(3075,1774)(3975,1774)
\path(3075,1774)(3975,1774)
\path(4920,1774)(5550,1774)
\path(4920,1774)(5550,1774)
\path(825,3124)(600,4834)
\path(825,3124)(600,4834)
\blacken\path(645.398,4718.939)(600.000,4834.000)(585.911,4711.112)(645.398,4718.939)
\path(825,3124)(1050,5239)
\path(825,3124)(1050,5239)
\blacken\path(1067.137,5116.500)(1050.000,5239.000)(1007.474,5122.847)(1067.137,5116.500)
\path(600,4924)(1230,3619)
\path(600,4924)(1230,3619)
\blacken\path(1150.814,3714.024)(1230.000,3619.000)(1204.847,3740.109)(1150.814,3714.024)
\path(1275,3574)(1095,5284)
\path(1275,3574)(1095,5284)
\blacken\path(1137.397,5167.800)(1095.000,5284.000)(1077.727,5161.519)(1137.397,5167.800)
\path(1725,4024)(690,4834)
\path(1725,4024)(690,4834)
\blacken\path(802.990,4783.668)(690.000,4834.000)(766.011,4736.418)(802.990,4783.668)
\path(1770,4024)(1095,5374)
\path(1770,4024)(1095,5374)
\blacken\path(1175.498,5280.085)(1095.000,5374.000)(1121.833,5253.252)(1175.498,5280.085)
\path(600,4924)(2085,4474)
\path(600,4924)(2085,4474)
\blacken\path(1961.457,4480.090)(2085.000,4474.000)(1978.857,4537.512)(1961.457,4480.090)
\path(2175,4474)(1140,5329)
\path(2175,4474)(1140,5329)
\blacken\path(1251.622,5275.703)(1140.000,5329.000)(1213.409,5229.445)(1251.622,5275.703)
\path(645,4924)(2535,4924)
\path(645,4924)(2535,4924)
\blacken\path(2415.000,4894.000)(2535.000,4924.000)(2415.000,4954.000)(2415.000,4894.000)
\path(3120,5374)(645,4969)
\path(3120,5374)(645,4969)
\blacken\path(758.580,5017.985)(645.000,4969.000)(768.270,4958.772)(758.580,5017.985)
\path(3120,5374)(1140,5374)
\path(3120,5374)(1140,5374)
\blacken\path(1260.000,5404.000)(1140.000,5374.000)(1260.000,5344.000)(1260.000,5404.000)
\path(2670,4924)(1275,5284)
\path(2670,4924)(1275,5284)
\blacken\path(1398.690,5283.063)(1275.000,5284.000)(1383.697,5224.966)(1398.690,5283.063)
\path(3975,5374)(5910,5554)
\path(3975,5374)(5910,5554)
\blacken\path(5793.295,5513.014)(5910.000,5554.000)(5787.737,5572.756)(5793.295,5513.014)
\path(4020,5374)(6360,5149)
\path(4020,5374)(6360,5149)
\blacken\path(6237.680,5130.623)(6360.000,5149.000)(6243.422,5190.348)(6237.680,5130.623)
\path(6045,5599)(4470,4924)
\path(6045,5599)(4470,4924)
\blacken\path(4568.480,4998.845)(4470.000,4924.000)(4592.115,4943.696)(4568.480,4998.845)
\path(4425,4879)(6315,5059)
\path(4425,4879)(6315,5059)
\blacken\path(6198.385,5017.758)(6315.000,5059.000)(6192.696,5077.488)(6198.385,5017.758)
\path(6000,5599)(4920,4519)
\path(6000,5599)(4920,4519)
\blacken\path(4983.640,4625.066)(4920.000,4519.000)(5026.066,4582.640)(4983.640,4625.066)
\path(5325,4024)(6000,5554)
\path(5325,4024)(6000,5554)
\blacken\path(5979.011,5432.101)(6000.000,5554.000)(5924.116,5456.319)(5979.011,5432.101)
\path(4920,4474)(6360,5014)
\path(4920,4474)(6360,5014)
\blacken\path(6258.174,4943.775)(6360.000,5014.000)(6237.107,4999.955)(6258.174,4943.775)
\path(5370,4024)(6405,5014)
\path(5370,4024)(6405,5014)
\blacken\path(6339.020,4909.374)(6405.000,5014.000)(6297.546,4952.733)(6339.020,4909.374)
\path(6000,5599)(5775,3619)
\path(6000,5599)(5775,3619)
\blacken\path(5758.741,3741.620)(5775.000,3619.000)(5818.357,3734.845)(5758.741,3741.620)
\path(5775,3529)(6450,5059)
\path(5775,3529)(6450,5059)
\blacken\path(6429.011,4937.101)(6450.000,5059.000)(6374.116,4961.319)(6429.011,4937.101)
\path(6270,3079)(6495,5104)
\path(6270,3079)(6495,5104)
\blacken\path(6511.565,4981.421)(6495.000,5104.000)(6451.932,4988.047)(6511.565,4981.421)
\path(6270,3124)(6045,5464)
\path(6270,3124)(6045,5464)
\blacken\path(6086.348,5347.422)(6045.000,5464.000)(6026.623,5341.680)(6086.348,5347.422)
\path(5595,1774)(3975,694)
\path(5595,1774)(3975,694)
\blacken\path(4058.205,785.526)(3975.000,694.000)(4091.487,735.603)(4058.205,785.526)
\path(3975,649)(4830,1684)
\path(3975,649)(4830,1684)
\blacken\path(4776.703,1572.378)(4830.000,1684.000)(4730.445,1610.591)(4776.703,1572.378)
\path(4875,1729)(3120,739)
\path(4875,1729)(3120,739)
\blacken\path(3209.778,824.088)(3120.000,739.000)(3239.257,771.829)(3209.778,824.088)
\path(5550,1774)(3165,649)
\path(5550,1774)(3165,649)
\blacken\path(3260.733,727.327)(3165.000,649.000)(3286.330,673.061)(3260.733,727.327)
\path(3975,649)(3975,1729)
\path(3975,649)(3975,1729)
\blacken\path(4005.000,1609.000)(3975.000,1729.000)(3945.000,1609.000)(4005.000,1609.000)
\path(4020,1774)(3120,784)
\path(4020,1774)(3120,784)
\blacken\path(3178.523,892.973)(3120.000,784.000)(3222.919,852.613)(3178.523,892.973)
\path(3030,1819)(3975,694)
\path(3030,1819)(3975,694)
\blacken\path(3874.846,766.589)(3975.000,694.000)(3920.788,805.180)(3874.846,766.589)
\path(3075,1774)(3075,784)
\path(3075,1774)(3075,784)
\blacken\path(3045.000,904.000)(3075.000,784.000)(3105.000,904.000)(3045.000,904.000)
\path(1275,1729)(2985,694)
\path(1275,1729)(2985,694)
\blacken\path(2866.806,730.471)(2985.000,694.000)(2897.874,781.801)(2866.806,730.471)
\path(1275,1774)(3840,649)
\path(1275,1774)(3840,649)
\blacken\path(3718.056,669.726)(3840.000,649.000)(3742.155,724.673)(3718.056,669.726)
\path(3975,649)(2175,1684)
\path(3975,649)(2175,1684)
\blacken\path(2293.983,1650.191)(2175.000,1684.000)(2264.075,1598.176)(2293.983,1650.191)
\path(2130,1774)(2985,739)
\path(2130,1774)(2985,739)
\blacken\path(2885.445,812.409)(2985.000,739.000)(2931.703,850.622)(2885.445,812.409)
\path(4425,4879)(3975,1819)
\path(4425,4879)(3975,1819)
\blacken\path(3962.779,1942.088)(3975.000,1819.000)(4022.140,1933.358)(3962.779,1942.088)
\path(3075,1819)(870,3124)
\path(3075,1819)(870,3124)
\blacken\path(988.549,3088.699)(870.000,3124.000)(957.990,3037.064)(988.549,3088.699)
\path(5325,4024)(5550,1864)
\path(5325,4024)(5550,1864)
\blacken\path(5507.729,1980.246)(5550.000,1864.000)(5567.406,1986.462)(5507.729,1980.246)
\path(4920,1774)(2670,4879)
\path(4920,1774)(2670,4879)
\blacken\path(2764.706,4799.433)(2670.000,4879.000)(2716.121,4764.227)(2764.706,4799.433)
\path(3075,5374)(4830,4474)
\path(3075,5374)(4830,4474)
\blacken\path(4709.532,4502.063)(4830.000,4474.000)(4736.911,4555.453)(4709.532,4502.063)
\path(2175,4429)(5730,3574)
\path(2175,4429)(5730,3574)
\blacken\path(5606.312,3572.892)(5730.000,3574.000)(5620.342,3631.229)(5606.312,3572.892)
\path(6270,3124)(2175,1819)
\path(6270,3124)(2175,1819)
\blacken\path(2280.226,1884.020)(2175.000,1819.000)(2298.444,1826.853)(2280.226,1884.020)
\path(1275,1774)(1725,3979)
\path(1275,1774)(1725,3979)
\blacken\path(1730.399,3855.425)(1725.000,3979.000)(1671.611,3867.422)(1730.399,3855.425)
\path(1298,3562)(3953,5272)
\path(1298,3562)(3953,5272)
\blacken\path(3868.359,5181.801)(3953.000,5272.000)(3835.870,5232.244)(3868.359,5181.801)
\path(1725,4024)(2175,4474)
\path(1725,4024)(2175,4474)
\put(15,4924){\makebox(0,0)[lb]{\smash{{$\alpha(x_1)$}}}}
\put(870,5689){\makebox(0,0)[lb]{\smash{{$\beta(x_1)$}}}}
\put(6135,5824){\makebox(0,0)[lb]{\smash{{$\alpha(x_2)$}}}}
\put(6585,5104){\makebox(0,0)[lb]{\smash{{$\beta(x_2)$}}}}
\put(3930,64){\makebox(0,0)[lb]{\smash{{$\alpha(x_3)$}}}}
\put(2715,109){\makebox(0,0)[lb]{\smash{{$\beta(x_3)$}}}}
\put(1185,3214){\makebox(0,0)[lb]{\smash{{$x_1$}}}}
\put(2040,4069){\makebox(0,0)[lb]{\smash{{$x_1$}}}}
\put(2895,4924){\makebox(0,0)[lb]{\smash{{$\bar{x}_1$}}}}
\put(3930,4924){\makebox(0,0)[lb]{\smash{{$x_2$}}}}
\put(4920,4024){\makebox(0,0)[lb]{\smash{{$\bar{x}_2$}}}}
\put(5865,3214){\makebox(0,0)[lb]{\smash{{$\bar{x}_2$}}}}
\put(5100,1954){\makebox(0,0)[lb]{\smash{{$x_3$}}}}
\put(3480,1954){\makebox(0,0)[lb]{\smash{{$\bar{x}_3$}}}}
\put(1635,1909){\makebox(0,0)[lb]{\smash{{$\bar{x}_3$}}}}
\end{picture}
}

\end{center}
\caption{Part of the digraph $H'(\cal F)$ when ${\cal F}=(x_1\vee x_2\vee \bar{x}_3)\wedge{}(\bar{x}_1\vee{}\bar{x}_2\vee{}x_3)\wedge{}(x_1\vee{}\bar{x}_2\vee\bar{x}_3)$. For better readability the vertices $c_1,c_2,c_3$ are not shown.}\label{}
\end{figure}
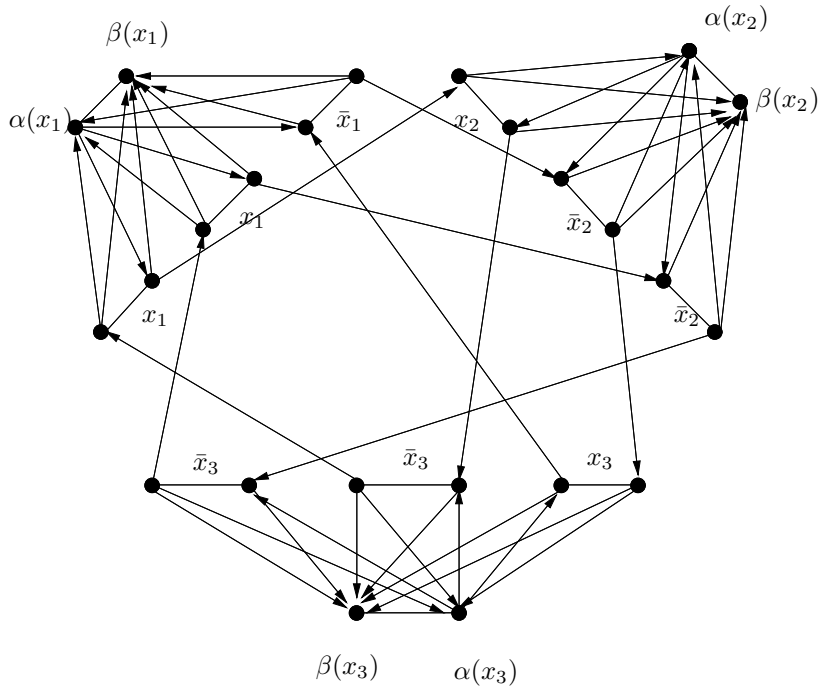

By Lemma \ref{excellentLTT} it suffices to show that $H$ has an excellent ordering if and only if $\cal F$ is satisfiable.

First suppose that $H$ has an excellent ordering. By Lemma \ref{excellentLTT} this means that the partially oriented complete graph $H^c$ has a completion $T$ as a locally transitive tournament. We claim that the following is a satisfying truth assignment: If the edge $\alpha{}(x_i)\beta{}(x_i)$ is oriented in $T$  as $(\alpha{}(x_i),\beta{}(x_i))$ then let $x_i=False$ and if it is oriented as $(\beta{}(x_i),\alpha{}(x_i))$ then let $x_i=True$. First observe that, by Proposition \ref{fixvar}, this implies that  for each $i\in [n]$ the variable  $x_i$ is false if and only if each of the edges $a_{i,j}b_{i,j}$, $j\in [p_i]$ are oriented as 
$(a_{i,j},b_{i,j})$ and each of the edges $u_{i,r}v_{i,r}$, $r\in [q_i]$ are oriented as $(v_{i,r},u_{i,r})$.

We now use this to show that each of the clauses of $\cal F$ are satisfied by our truth assignment.
As $T$ is locally transitive, for each of the induced subdigraphs 
$\induce{T}{W_j}$, $j\in [m]$ the out-neighbourhood of $c_j$ is acyclic which implies that at least one of three arcs of $H$ which correspond to the literals of $\cal F$ is oriented as $(c_{j2},c_{j1})$. If this arc corresponds to the literal $x_s$ then, by the identification rule above this is an  arc of the form 
$(b_{s,t},a_{s,t})$ so the variable $x_s$ is true and $C_j$ is satisfied. If the arc corresponds to the literal $\bar{x}_s$ then the identification rule implies that this is an arc of the form $(v_{s,t},u_{s,t})$, implying that  $\bar{x}_s$ is true so again $C_j$ is satisfied. Thus we have shown that $\cal F$ is satisfiable if $H^c$ has a locally transitive completion ($H$ has an excellent ordering). 

Now suppose that $t:\{x_1,\ldots{},x_n\}\rightarrow \{True,False\}$ is a satisfying truth assignment for $\cal F$. We shall use this truth assignment to construct an excellent ordering of the  pog $H'$. Recall that this  is also an excellent ordering of the directed part $H$ of $H'$.

We first orient the edges $\alpha(x_1)\beta(x_1),\ldots{}\alpha(x_n)\beta(x_n)$ as follows: If $x_i=True$ then orient $\alpha(x_i)\beta(x_i)$ as $(\beta(x_i),\alpha(x_i))$ and otherwise orient as $(\alpha(x_i),\beta(x_i))$. Denote by $\hat{H}$ the resulting pog. It follows from Proposition
\ref{fixvar}, the way we made identifications between vertices of the $W_j$'s and variable vertices and the fact that $t$ is a satisfying truth assignment that   we can now orient all the remaining  edges of $\hat{H}$  (recall that those correspond to the literals) uniquely so that the resulting full orientation $\stackrel{\rightarrow}{H}$ of $H'$ satisfies that the in- and out-neighbourhood of each vertex is still acyclic.

We now construct an excellent ordering for $\stackrel{\rightarrow}{H}$.
Denote by $A(x_i)$ ($B(x_i)$), $i\in [n]$ the set of out-neighbours (in-neighbours) of $\alpha(x_i)$ in $\stackrel{\rightarrow}{H}$. Note that if $t(x_i)=False$, then 
we must have $A(x_i)=\{b_{i,1},\ldots{},b_{i,p_i},u_{i,1},\ldots{},u_{i,q_i},\beta(x_i)\}$,  $B(x_i)=\{a_{i,1},\ldots{},a_{i,p_i},v_{i,1},\ldots{},v_{i,q_i}\}$ and there is no oriented arc from $B(x_i)$ to $A(x_i)$. Similarly, if $t(x_i)=True$, then $A(x_i)=\{b_{i,1},\ldots{},b_{i,p_i},u_{i,1},\ldots{},u_{i,q_i}\}$, $B(x_i)=\{a_{i,1},\ldots{},a_{i,p_i},v_{i,1},\ldots{},v_{i,q_i},\beta(x_i)\}$ and there is no oriented arc from $A(x_i)$ to $B(x_i)$. 

Furthermore observe that $\beta(x_i)$ has no out-neighbour when $t(x_i)=False$ and precisely one out-neighbour, namely $\alpha(x_i)$ when $t(x_i)=True$.
Let $1\leq i_1<i_2<\ldots{}<i_k\leq n$ and $1<j_1<j_2<\ldots{}<j_g\leq n$ denote the indices of the true, respectively the false variables.
Consider the following 
cyclic ordering  $\cal O$ of $V(\stackrel{\rightarrow}{H})$:

$\alpha(x_{i_1}),\alpha(x_{i_2}),\ldots{},\alpha(x_{i_k}),c_1,c_2,\ldots{},c_m,
A(x_{i_1}),\ldots{},A(x_{i_k}),B(x_{j_1}),\ldots{},B(x_{j_g}),\alpha(x_{j_1}),\ldots{},\alpha(x_{j_g}),$\\
 $A(x_{j_1}),\ldots{},A(x_{j_g}), B(x_{i_1}),\ldots{},B(x_{i_k}),\alpha(x_{i_1})$,\\
\noindent{} where the ordering inside each $A(x_i),B(x_i)$ is as according to the way we listed those sets above.

We shall prove that the ordering $\cal O$ is excellent.
Suppose for contradiction that there is a pair of arcs $(v_i,v_j)$ and $(v_s,v_t)$ with the vertices occurring in the order $v_i,v_t,v_s,v_j$ according to $\cal O$.

\begin{itemize}
\item We cannot have $v_i=\alpha(x_{i_f})$ for some $f\in [k]$ because there is no backward arc in the interval  of $\cal O$ from $\alpha(x_{i_f})$ to (the end of) $A(x_f)$ ($\alpha(x_{i_f})$ is only adjacent to vertices in $A(x_{i_f})$). Similarly, we cannot have $v_i$ in the interval $[\alpha(x_{j_1}),\alpha(x_{j_g})]$.
\item We cannot have $v_i=c_p$ for some $p\in [m]$ because the only arcs incident to $c_p$ are from $c_p$ to the six vertices  which correspond to its three litterals and we ordered the $A$ and $B$ sets and $\alpha(x_{j_1}),\ldots{},\alpha(x_{j_g})$ in such a way that any arc between them goes forward in the ordering. In particular there are no backwards arcs with respect to the ordering in the interval
$$A(x_{i_1}),\ldots{},A(x_{i_k}),B(x_{j_1}),\ldots{},B(x_{j_g}),\alpha(x_{j_1}),\ldots{},\alpha(x_{j_g}),A(x_{j_1}),\ldots{},A(x_{j_g}), B(x_{i_1}),\ldots{},B(x_{i_k})$$
\item We cannot have $v_i$ in the interval  $A(x_{i_1}),\ldots{},A(x_{i_k})$ since all out-neighbours of those vertices are in the interval $B(x_{i_1}),\ldots{},B(x_{i_k})$ and then  the  remark above implies the claim. Similarly, we cannot have $v_i$ in the interval $A(x_{j_1}),\ldots{},A(x_{j_g})$.
\item We cannot have $v_i$ in the interval $B(x_{j_1}),\ldots{},B(x_{j_g})$ because there are no backward arcs in the interval $B(x_{j_1}),\ldots{},B(x_{j_g}),\alpha(x_{j_1}),\ldots{},\alpha(x_{j_g}),
 A(x_{j_1}),\ldots{},A(x_{j_g})$ and this contains all out-neighbours of such a $v_i$.
\item Finally we cannot have $v_i$ in the interval $B(x_{i_1}),\ldots{},B(x_{i_k})$ because all arcs out of a vertex in this interval remains inside the interval 
$B(x_{i_1}),\ldots{},B(x_{i_k}),\alpha(x_{i_1}),\alpha(x_{i_2}),\ldots{},\alpha(x_{i_k})$ and there is no backward arc here.
\end{itemize}

Thus we have shown that $\cal O$ is excellent and hence, by Lemma \ref{excellentLTT} the partially oriented complete graph $H^c$ has a completion to a locally transitive tournament.
 \qed

\section{Polynomial cases}
\label{polys}

Let $G = (V,E)$ be a graph. The {\bf auxiliary graph} $G^+$ of $G$ is defined 
as follows. The vertex set of $G^+$ consists of all ordered pairs $(u,v)$ for 
all $uv \in E$ (note that every edge of $G$ gives rise to two vertices of 
$G^+$).
Two vertices $(u,v)$ and $(u',v')$ of $G^+$ are adjacent if and only if 
one of the following conditions holds:

\begin{itemize}
\item $u = u'$ and $vv' \notin E$;
\item $uu' \notin E$ and $v = v'$;
\item $u = v'$ and $v = u'$.
\end{itemize}

\begin{lemma} \cite{hellJGT20} \label{auxiliary}
A graph $G$ is local tournament orientable if and only if $G^+$ is bipartite.
Moreover, when $G^+$ is bipartite, for any two vertices $(u,v), (u',v')$ of 
odd distance in $G^+$, a local tournament of $G$ must contain
exactly one of them as an arc. In particular, the arcs of every local tournament 
orientation of $G$ correspond to a colour class of $G^+$.
\qed
\end{lemma}

\begin{theorem} 
The orientation completion problem is polynomial time solvable for 
the class of local tournaments is polynomial time solvable.
\end{theorem}

\pf Let $P = (V, A\cup E)$ be a partially oriented graph and let $G = UG(P)$.
The arc set $A$ corresponds to a subset $S$ of the vertex set of $G^+$. 
According to Lemma \ref{auxiliary} $P$ can be completed to a local tournament 
if and only if $G^+$ is bipartite and $S$ is contained in a colour class of $G^+$.
Checking whether $G^+$ is bipartite and in the case when $G^+$ is bipartite 
whether $S$ is contained in a colour class of $G^+$ can be done 
in polynomial time. 
\qed

An oriented graph $D = (V,A)$ is called {\bf quasi-transitive} if for any 
three vertices $x, y, z$, $(x,y) \in A$ and $(y,z) \in A$ together imply
there is an arc between $x$ and $z$ in either direction, cf. \cite{bangJGT20b}.  
In a similar way we can define an auxiliary graph for each graph $G$
which can be used to determine whether $G$ can be completed to 
a quasi-transitive oriented graph, cf. \cite{hellJGT20}. 
This also implies that the orientation completion problem for the class of 
quasi-transitive oriented graphs is solvable in polynomial time.

We next consider the orientation completion problem for acyclic local tournaments.
By Theorem \ref{ship}, a partially oriented graph $P$ can be completed to 
an acyclic local tournament if and only if $UG(P)$ is a proper interval graph.
Since every (proper) interval graph is chordal, it has 
a perfect elimination ordering, which is a vertex ordering $\prec$ such that
if $x \prec y \prec z$ and $xy, xz$ are edges then $yz$ is an edge. 

There is a simple algorithm which determines whether a graph is a proper 
interval graph and completes it to an acyclic local tournament (orientation) 
if it is, cf. \cite{hellJGT20}.  
Let $G$ be a graph and let $\prec$ be a vertex ordering of $G$. For two
ordered pairs $(u,v), (u',v')$ of vertices of $G$, we say that $(u,v)$ is 
{\bf lexicographically smaller than $(u',v')$ with respect to $\prec$}
if either $u \prec u'$ or $u = u'$ and $v \prec v'$. The following 
algorithm is taken from \cite{hellJGT20}.

\medskip

{\bf Lexicographic 2-Colouring}: Let $G = (V,E)$ be a graph which is 
chordal and whose auxiliary graph $G^+$ is bipartite.

\begin{enumerate}
\item Find a perfect elimination ordering $\prec$ of $G$.
\item While there exist uncoloured vertices in $G^+$: colour 
      lexicographically the smallest uncoloured vertex $(u,v)$ red and extend
      it a red/blue colouring of the connected component of $G^+$ containing 
      $(u,v)$.
\end{enumerate}

\begin{lemma} \cite{hellJGT20} \label{slexico}
Let $G = (V,E)$ be a graph which is chordal and whose auxiliary graph $G^+$ is 
bipartite. Let $R$ be the set of all red vertices of $G^+$ obtained by
the Lexicographic 2-Colouring algorithm above. Then $G$ is a proper  
interval graph if and only if $(V,R)$ is an acyclic local tournament 
orientation of $G$.
\qed
\end{lemma}

Rose, Tarjan and Lueker \cite{rtl} developed a linear time algorithm for finding 
a perfect elimination ordering in a chordal graph. The algorithm is called 
the {\bf Lexicographic Breadth First Search} (LBFS) which is a refinement of 
the classical {\bf Breadth First Search} for graphs. Begining with 
an arbitrary vertex of the graph, LBFS always labels the next vertex to be 
one whose 
neighbourhood among the labeled vertices is the lexicographically largest
(that is, if $v_n, v_{n-1}, \dots, v_{i+1}$ are labeled, then the next 
labeled vertex $v_i$ for which 
$\{j:\ j > i\ \mbox{and}\ v_iv_j\ \mbox{is\ an\ edge}\}$ 
is the lexicographically largest among all unlabled vertices.
(For two distinct subsets $S, T \subseteq \{v_n, v_{n-1}, \dots, v_{i-1}\}$,
$S$ is {\bf lexicographically larger} than $T$ if the vertex in
$S \triangle T$ with the largest subscript belongs to $S$.)

Suppose that $G$ is local tournament orientable. By Lemma \ref{auxiliary}, 
if $D$ is a local tournament orientation $D$ of $G$, then $D$

\begin{itemize}
\item does not contain two arcs whose corresponding vertices are of 
      odd distance in $G^+$, and
\item does not contain exactly one of any two arcs whose corresponding vertices 
      are of even distance in $G^+$.
\end{itemize}

A partial orientation of $G$ is called {\bf consentaneous} if it satisfies
the two properties listed above. 

\begin{theorem} \label{acyclic-ext1}
Let $P = (V, A\cup E)$ be a partially oriented graph. Suppose that $UG(P)$ is 
a proper interval graph and $P$ is consentaneous. Then $P$ can be 
completed to an acyclic local tournament if and only if $P$ does not 
contain a directed cycle.
\end{theorem}

\pf If $P$ contains a directed cycle then it cannot be completed to an
acyclic oriented graph and hence not to an acyclic local tournament.
For the other direction, we first show that $P$ admits a perfect elimination 
ordering $v_1, v_2, \dots, v_n$ such that all arcs are forward, that is,
if $(v_i,v_j)$ is an arc then $i < j$. To obtain such an ordering we
apply a modified LBFS begining with a vertex of outdegree 0, with preferences 
(in the case of ties) 
given to vertices having no out-neighbours among unlabeled vertices.

Let $v_1, v_2, \dots, v_n$ be an ordering obtained by the modified LBFS.
According to \cite{rtl}, it is a perfect elimination ordering. Suppose that 
the ordering contains backward arc. Let $(v_i,v_j) \in A$ be 
a backward arc having the largest subscript $i$. Since $(v_i,v_j)$ is 
backward, we have $i > j$. The choice of $v_n$ implies $n > i$.
Since $i > j$, at the time of labeling $v_i$ the vertex $v_j$ is an unlabled 
out-neighbour of $v_i$. The LBFS rule ensures that $v_i$ is a vertex having
the lexicographically largest neighbourhood among the vertices 
$v_n, \dots, v_{i+1}$. If the neighbourhood of $v_i$ (among the labeled vertices)
is lexicographically larger than 
the neighbourhood of $v_j$, some some vertex $v_{\ell}$ with $\ell > i$
adjacent to $v_i$ but not to $v_j$ in $P$. The assumption that $P$ is 
consentaneous implies $(v_{\ell},v_i)$ is an arc which is backward 
with respect to the ordering. This contradicts the choice of $(v_i,v_j)$. 
Hence $v_i$ and $v_j$ must have the same neighbourhood among 
the labeled vertices. But then 
the rule prefers $v_j$ to $v_i$ for the next labeled vertex, unless $v_j$ 
has an out-neighbour $v_k$ among unlabeled vertices. A similar proof above 
(when applied to $v_j, v_k$) implies  $v_j$ and $v_k$ must have the same 
neighbourhood among the labeled vertices. Continuing this way, we obtain
a directed cycle (consisting of vertices with the lexicographically largest
neighbourhood), which contradicts the assumption. 
Hence $v_1, v_2, \dots, v_n$ is a perfect elimination ordering of $P$ 
that contains no backward arcs.

Now we apply the Lexicographic 2-Colouring algorithm using the perfect
elimination ordering to obtain a red/blue colouring of the vertices
of $UG(P)^+$. Let $R$ be the set of red vertices of $UG(P)^+$ produced
by the algorithm. Since the perfect elimination ordering has no backward
arc from $A$, $A \subseteq R$. Hence by Lemma \ref{slexico}, $(V,R)$ 
is an acyclic local tournament which is an orientation completion of $P$.
\qed

\begin{corollary} \label{acyclic-ext2}
The orientation completion problem for the class of acyclic local tournaments
is solvable in polynomial time.
\end{corollary}

\pf Suppose that a partially oriented graph $P = (V,A \cup E)$ is given.
Denote $G = UG(P)$. If $G^+$ is not bipartite, then the answer is "no". 
Obtain the minimal consentaneous partial oriented graph $P' = (V, A' \cup E')$
from $P$ by orienting (if needed) some edges in $E$. If $P'$ contains 
a directed cycle, then the answer is again "no" by Lemma \ref{auxiliary} and 
Theorem \ref{acyclic-ext1}. Otherwise, $P'$ contains no directed cycle and 
we can complete $P'$ to an acyclic local tournament orientation of $P'$
according to Theorem \ref{acyclic-ext1}. This acyclic local tournament is
also a completion of $P$. All these steps can be done in polynomial time.
\qed

\begin{corollary} \cite{kkk}
The problem of extending partial proper interval representations of
proper interval graphs is solvable in polynomial time.
\end{corollary}

\pf We show how to reduce the problem of extending partial proper interval 
representations of proper interval graphs to the orientation completion problem 
for the class of acyclic local tournaments which is polynomial time solvable
according to Corollary \ref{acyclic-ext2}. 
Suppose that $G$ is a proper interval graph and $H$ is an induced subgraph 
of $G$. Given a proper interval representation $I_v, v \in V(H)$ of $H$ 
(i.e., a partial proper interval representation of $G$), we obtain an
orientation of $H$ in such a way that $(u,v)$ is an arc if and only if
$I_u$ contains the left endpoint of $I_v$. The oriented edges together with
the remaining edges in $G$ yield a partial orientation of $G$. 
This partial orientation of $G$ can be completed to an acyclic local 
tournament if and only if the partial representation of $H$ can be extended
to a proper interval representation of $G$. 
\qed

\medskip

We now return to the orientation completion problem for local transitive
tournaments. This problem is NP-complete in general and remains so even
for complete graphs as shown in Section~\ref{npc}.
We will show that the problem becomes polynomial time solvable if 
the input partially oriented graphs are all friendly (see definition 
below).

Let $P = (V, A \cup E)$ be a partially oriented graph and let $G = UG(P)$. 
A triple of vertices $x, y, z$ in $P$ is called {\bf bad} if 
\begin{itemize}  
\item $xyzx$ is a triangle in $G$,
\item the three edges in $xyzx$ correspond to vertices from
      three different connected components in $G^+$, and
\item exactly two edges in $xyzx$ are oriented in $P$.
\end{itemize} 

If $P$ is consentaneous and has no bad triple then it is called {\bf friendly}.
When $G$ is a complete graph, the three edges of each fixed triangle 
correspond to vertices from three connected components of $G^+$.
Thus if a partially oriented complete graph is friendly then 
the arcs induce vertex disjoint tournaments. 

\begin{lemma} \label{merge}
Let $P = (V,A \cup E)$ be a partially oriented complete graph.
Suppose that the arc set $A$ induces two vertex disjoint tournaments $T'$ and
$T''$ with $V(T') \cup V(T'') = V$ which are both locally transitive. 
Then $P$ can be completed to a tournament that is also locally transitive.
\end{lemma}

\pf Since $T'$ is locally transitive, by Theorem \ref{moon} $T'$ is obtained
from a highly regular tournament $H'$ with round ordering 
$u_0, u_1, \dots, u_{2a}$ by substituting transtive tournament $T_i$ 
for $u_i$ for each $i = 0, 1, \dots, 2a$, that is,
$T' = H'[X_0, X_1, \dots, X_{2a}]$.
Similarly we have $T'' = H''[Y_0, Y_1, \dots, Y_{2b}]$
where $H''$ is a highly regular tournament with round ordering 
$v_0, v_1, \dots, 2b$ and each $Y_i$ is a transitive tournament replacing 
$v_i$. Without loss of generality assume $a \geq b$.
Let $T = H'[(X_0 \cup Y_0), \dots, (X_b \cup Y_b), X_{b+1}, \dots, X_a, 
(X_{a+1} \cup Y_{b+1}), \dots, (X_{a+b} \cup Y_{2b}), X_{a+b+1}, \dots, X_{2a}]$.
It is easy to verify that the tournament $T$ is an orientation completion of $P$
and is locally transitive.
\qed

The following theorem characterizes friendly partially oriented complete graphs 
which can be completed to a tournament that is locally transitive.

\begin{theorem} \label{friendly-complete} 
Let $P = (V,A \cup E)$ be a friendly partially oriented complete graph.
Then $P$ can be completed to a tournament that is locally transitive 
if and only if no directed triangle is contained in the in-neighbourhood or 
the out-neighbourhood of any vertex in $P$.
\end{theorem}

\pf We only prove the sufficiency as the necessity is obvious.
Suppose that $P$ has no directed triangle contained in 
the in-neighbourhood or the out-neighbourhood of any vertex. 
Since it is friendly, $V$ can be partitioned into vertex disjoint tournaments 
such that no arc is between any two of them. Since no directed cycle is 
contained in the in-neighbourhood or the out-neighbourhood of any vertex, 
each tournament is locally transitive. By Lemma \ref{merge}, any two 
such tournaments can be completed to a tournament that is locally transitive and 
therefore $P$ can be completed to a tournament that is locally transitive.
\qed

By Theorem \ref{friendly-complete}, to determine whether a friendly partially 
oriented complete graph can be completed to a tournament that is locally 
transitive, one only needs to check if it has a directed triangle contained 
in the in-neighbourhood or the out-neighbourhood of a vertex. In the case
when no such a triangle exists, following the proof of 
Theorem \ref{friendly-complete} we can complete the given partially
oriented graph to a locally transitive tournament.
All these can be done in polynomial time. Hence we have the following:

\begin{corollary} 
The problem of deciding whether a friendly partially oriented complete graph 
can be completed to a tournament that is locally transitive and constructing 
such an orientation completion (if one exists) is solvable in polynomial 
time.  
\qed
\end{corollary}

Let $P = (V, A \cup E)$ be a partially oriented graph and $G = UG(P)$. 
Two vertices in $P$ are {\bf similar} if they have the same closed neighbourhood 
in $G$. An edge $xy$ (oriented or not) in $G$ is called {\bf balanced} if $x$ 
and $y$ are similar and is called {\bf unbalanced} otherwise.  
A {\bf cell} of $P$ is a maximal subgraph of pairwise similar vertices. 
Note that each cell is a complete subgraph and two cells are either completely 
adjacent or completely nonadjacent. 
A vertex of $P$ is {\bf universal} if it is adjacent to all other vertices.
Clearly universal vertices are similar to each other. The cell on the
universal vertices of $G$ will be called {\bf universal} and any other cell 
will be called {\bf non-universal}. Each edge $xy$ in a cell corresponds to 
a connected component of $G^+$ on the two vertices $(x,y), (y,x)$.
We shall call such a component of $G^+$ a {\bf thin} component and others
{\bf thick} components. 

We recall some results from \cite{huangJCT63} which are described in
the next two theorems and will be useful in the discussion.

\begin{theorem} \cite{huangJCT63} \label{cells}
Let $D$ be a local tournament. Then the following statements hold:

\begin{itemize}
\item If $B$ is a non-universal cell then $B$ induces a transitive tournament.
\item If $B$ is a non-universal cell and $v \notin B$ is adjacent to the 
      vertices in $B$, then $v$ either completely dominates $B$ or is 
      completely dominated by $B$. 
\item If $B$ and $B'$ are two adjacent non-universal cells, then either $B$ 
      completely dominates $B'$ or is completely dominated by $B'$.
\end{itemize}
\qed
\end{theorem}

\begin{theorem} \cite{huangJCT63} \label{implication-class} 
Let $G$ be a connected graph that is local tournament orientable and let 
$C_1, C_2, \dots, C_k$ be the connected components of the complement 
$\overline{G}$ of $G$. Then the following statements hold. 

\begin{itemize}
\item Suppose that $\overline{G}$ is not bipartite. Then $\overline{G}$ has 
      exactly one connected component (i.e., $k = 1$) and $G^+$ has exactly 
      one thick component. Each cell of $G$ is non-universal. 
      A local tournament orientation of $G$ is locally transitive if and only if 
      each cell is a transitive tournament.
\item Suppose that $\overline{G}$ is bipartite. Then the vertices in each fixed
      thick component of $G^+$ correspond to either all unbalanced edges of $G$
      within a fixed $C_i$ or all edges between two fixed $C_i$ and $C_j$ 
      ($i \neq j$). In the case when $k = 1$ and $G$ has at least two 
      vertices, each cell is non-universal, $G^+$ has exactly one thick 
      component, and moreover a local tournament orientation of $G$ is locally 
      transitive if and only if each cell is a transitive tournament.
\qed
\end{itemize}
\end{theorem}

\begin{theorem} \label{friendly-general}
Let $G$ be a connected graph that is local tournament orientable. 
Then a friendly partial orientation $P = (V, A\cup E)$ of $G$ can be
completed to a locally transitive local tournament if and only if 
$P$ has no directed cycle contained in
\begin{itemize} 
\item a non-universal cell, or
\item the in-neighbourhood or the out-neighbourhood of a vertex.
\end{itemize}
\end{theorem}

\pf The necessity follows from Theorem \ref{cells} and the fact that a locally
transitive local tournament has no directed cycle contained in 
the in-neighbourhood or the out-neighbourhood of any vertex.  

For the sufficiency suppose that $P$ has no directed cycle contained in
a non-universal cell, or in the in-neighbourhood or the out-neighbourhood of 
a vertex. In view of Theorem \ref{friendly-complete}, we may assume that 
$G$ is not complete. This implies that $\overline{G}$ has at least one 
non-trivial component and $G^+$ has at least one thick component. 

We first explain how to complete the orientation of each non-universal cell 
in $P$. By assumption no directed cycle is contained in any non-universal cell. 
Thus the orientation of each cell can be completed to
a transitive tournament. If $G^+$ has only one thick component then 
the resulting orientation of $G$ can be further completed to a local tournament 
orientation of $G$ that is locally transitive by Theorem \ref{implication-class}.
In particular, when $\overline{G}$ is not bipartite or $\overline{G}$ is 
bipartite and has exactly one connected component, the partial orientation of 
$G$ can be completed to a local tournament that is locally transitive. 
So we may assume that $\overline{G}$ is bipartite and has at least two components.
We may assume furthermore that each non-universal cell is 
a transitive tournament. 

Let $C_1, C_2, \dots, C_q$ be the connected components of $\overline{G}$ where
each $C_i$ with $1 \leq i \leq p$ is non-trivial and the rest are trivial 
(i.e., each consisting of a universal vertex of $G$). 
Consider a fixed $C_i$. According to Theorem \ref{implication-class}, 
all unbalanced edges of $G$ within $C_i$ correspond the vertices of 
a component of $G^+$. Since the partial orientation is friendly and hence 
consentaneous by definition, either all unbalanced edges of $G$ within $C_i$ 
are oriented or none of them is. 
In the latter case we orient all unbalanced edges of $G$ within $C_i$
using any one of the two colour classes of the component of $G^+$ 
corresponding the edges. Thus all edges of $G$ within $C_i$ are now oriented. 
The only edges in $G$ that are not oriented (if any) are between the components
$C_1, C_2, \dots, C_q$.

Arbitrarily choose a vertex $s_i \in C_i$ for each $i = 1, 2, \dots, q$ and
consider the (complete) subgraph $K$ of $G$ induced by $s_1, \dots, s_q$. 
Note that in each triangle of $K$ the three edges correspond to vertices from 
three components of $G^+$. Since $P$ is friendly, there cannot be exactly two 
edges in each triangle of $K$ are oriented in $P$.
This means that the subgraph of $P$ induced by $s_1, s_2, \dots, s_q$
is a friendly partial orientation of $K$.
By Theorem \ref{friendly-complete}, it can be completed to a full orientation
of $K$ that is locally transitive. Denote this tournament by $R$. 
We explain how $R$ can guide us to orient the remaining unoriented edges
(between the components).  

Let $(S_i,T_i)$ be the bipartition of $C_i$ for each $i = 1, 2, \dots, p$.
Without loss of generality assume $s_i \in S_i$ for each $i = 1, 2, \dots, p$.
Note that each $C_i$ with $p+1 \leq i \leq q$ consists of the single vertex
$s_i$. Suppose that $(s_i,s_j)$ is an arc in $R$. In the case when $(s_i,s_j)$ 
is in $P$, all edges between $C_i$ and $C_j$ are oriented since $P$
is consentaneous. So assume $(s_i,s_j)$ is not in $P$.  
If $1 \leq i, j \leq p$, then orient the edges between $C_i$ and $C_j$ in 
such a way that $S_i \dom S_j \dom T_i \dom T_j \dom S_i$; if 
$1 \leq i \leq p$ and $p+1 \leq j \leq q$, then 
$S_i \dom s_i \dom T_i$; if $1 \leq j \leq p$ and $p+1 \leq i \leq q$, then
$T_j \dom s_i \dom S_j$; if $p+1 \leq i, j \leq q$, then $s_i \dom s_j$.
We remark that in the case when $(s_i,s_j)$ is in $P$,
the edges between $C_i$ and $C_j$ are oriented in the same way
as defined above. 

Let $D$ denote the full orientation of $G$. We will show that $D$ is 
locally transitive. According to Theorem \ref{implication-class} the subdigraph 
$D_i$ of $D$ induced by $V(C_i)$ is locally transitive for each $i$. Hence
by Theorem \ref{roundLT} $D_i$ has a round ordering $u_1, u_2, \dots, u_t, u_1$.
We claim that $S_i$ induces a transitive tournament in $D_i$. 
Suppose to the contrary that the tournament induced by
$S_i$ contains a directed triangle. Without loss of generality assume
$u_1u_au_bu_1$ is such a triangle. Recall that $C_i$  is a connected component
in $\overline{G}$. In particular, each vertex of $D_i$ has a
a non-neighbour in $D_i$. Let $X = N_{D_i}^+(u_b) \cap N_{D_i}^-(u_a)$ and 
$Y = \{u_{a+1}, u_{a+2}, \dots, u_{b-1}\}$. 
Note that the vertices of $X$ appear consecutively in the round ordering, 
that is, $X = \{u_c, u_{c+1}, \dots, u_1, \dots, u_d\}$ for some $c, d$. 
Let $W = \{u_{b+1}, u_{b+2}, \dots, u_{c-1}\}$ and 
$Z = \{u_{d+1}, u_{d+2}, \dots, u_{a-1}\}$. 
The vertices in $W \cup X$ are pairwise adjacent as they are out-neighbours
of $u_b$. Similarly vertices in $Z \cup X$ are pairwise adjacent
as they are in-neighbours of $u_a$. Therefore 
the non-neighbours of each vertex in $X$ can only be in $Y$. 
In particular the non-neighbours of each vertex in $X \cap S_i$ are in 
$Y \cap T_i$ since any two vertices in $S_i$ are adjacent in $G$.
Since $u_a \in S_i$, it is adjacent to every vertex in $W \cap S_i$.
Every vertex in $W \cap S_i$ is adjacent to 
all vertices in $Y \cap T_i$ as they are out-neighbours of $u_a$. 
Similarly, every vertex in $Z \cap S_i$ is adjacent to all vertices in 
$Y \cap T_i$ as they are in-neighbours of $u_b$. 
Hence the non-neighbours of each vertex in $Y \cap T_i$ can only be in 
$X \cap S_i$. It follows that any path in $\overline{G}$ joining $u_1$ can only
have vertices in $(X \cap S_i) \cup (Y \cap T_i)$. 
Since $u_a \notin (X \cap S_i) \cup (Y \cap T_i)$, there is no path in
$\overline{G}$ joining $u_1$ and $u_a$, which contradicts
the fact that they are in the same connected component of $\overline{G}$.
Therefore $S_i$ must induce a transitive tournament in $D$. 
By symmetry, $T_i$ also induces a transitive tournament. 

Suppose to the contrary that $D$ has a directed triangle $xyz$ that is contained 
in the out-neighbourhood of $w$. The proof is similar if $D$ has a triangle
that is contained in the in-neighbourhood of a vertex. 
We claim each $D_i$ (the subdigraph induced by $V(C_i)$) contains 
at most one vertex from $\{w, x, y, z\}$. Suppose $D_i$ contains two vertices
(say $x, y$) from the triangle. If $x, y \in S_i$, then $z \in T_i$ because
every vertex not in $D_i$ either dominates both $x, y$ or is dominated by 
both of them and no vertex in $S_i$ forms a directed triangle with $x, y$.  
Since $w$ is dominates each of $x, y, z$, it can only be in $D_i$. Thus
all four vertices $w, x, y, z$ are in $D_i$, a contradiction to the fact
that $D_i$ is locally transitive. A similar argument shows that $T_i$ cannot
contain both $x, y$. By symmetry we may assume that $x \in S_i$ and $y \in T_i$. 
The above proof implies $z$ cannot be in $D_i$ (as otherwise either $z, x$ 
would be both in $S_i$ or $z, y$ would be both in $T_i$).  
Since $w$ dominates both $x, y$ and no vertex not in $D_i$ can have this 
property, $w$ is also in $D_i$. Since $z \notin D_i$ is dominated by both 
$w, y$, the vertex $w$ must be in $T_i$. Since $x$ is an in-neighbour of $y$,
it is adjacent to all in-neighbours of $y$ in $T_i$. 
Since $w$ dominates $y$ and $T_i$ induces a transitive tournament in $D$, 
each out-neighbour of $y$ is an out-neighbour of $w$.
Since $x$ is an out-neighbour of $w$, it is adjacent to all out-neighbours of
$w$ and in particular to all out-neighbours of $y$.  
Hence $x$ is adjacent to all other vertices in $D_i$. This contradicts the 
fact that $C_i$ is a connected component in $\overline{G}$. 
If $D_i$ contains $w$ and one of 
$x, y, z$ (say $x$). Then $w, x$ cannot be both in $S_i$ or both in $T_i$ 
as $z \notin D_i$ dominates $x$ and is dominated by $w$. On the other hand, 
since $w$ and $x$ both dominate $y \notin D_i$, they must be either both in $S_i$
or both in $T_i$. So $D_i$ cannot contain both $w$ and $x$. 
Therefore each $D_i$ contains at most one vertex from $\{w, x, y, z\}$.
Assume that $w \in D_{\alpha}, x \in D_{\beta}, y \in D_{\gamma}$ and 
$z \in D_{\ell}$. For each $i \in \{\alpha, \beta, \gamma, \ell\}$,
let $v_i$ be the only vertex in $V(D_i) \cap \{w,x,y,z\}$ (which may or 
may not be the vertex $s_i$ (defined above) that is also contained in $D_i$).
Denote by $F$ the subdigraph induced by $\{w,x,y,z\}$
($ = \{v_{\alpha},v_{\beta},v_{\gamma},v_{\ell}\}$).
Suppose that $(s_i,s_j)$ is an arc in $D$. Then $(v_i,v_j)$ is an arc
if either $v_i \in S_i$ and $v_j \in S_j$ or $v_i \in T_i$ and $v_j \in T_j$;
otherwise $(v_j,v_i)$ is an arc in $D$. 
It follows that the subdigraph induced by 
$\{s_{\alpha},v_{\beta},v_{\gamma},v_{\ell}\}$ is either the same as 
$F$ or is the same as the one obtained from $F$ by reversing the arcs between 
$v_{\alpha}$ and the other three vertices. In any case it is easy to
verify that the subdigraph induced by 
$\{s_{\alpha},v_{\beta},v_{\gamma},v_{\ell}\}$ 
has a directed triangle contained
in the in-neighbourhood of the fourth vertex or a directed triangle contained
in the out-neighbourhood of the fourth vertex. 
It follows that the subdigraph induced by
$s_{\alpha}, s_{\beta}, s_{\gamma}, s_{\ell}$ has a directed triangle contained
in the in-neighbourhood of the fourth vertex or a directed triangle contained
in the out-neighbourhood of the fourth vertex. This is is a contradiction
to the fact that the subdigraph induced by 
$s_{\alpha}, s_{\beta}, s_{\gamma}, s_{\ell}$ is locally transitive.
Therefore $D$ is a locally transitive local tournament which is an orientation
completion of the given partial orientation $P$ of $G$.
\qed

\begin{corollary} \label{friendly-ext} 
The problem of deciding whether a friendly partially oriented graph 
can be completed to a locally transitive local tournament and obtaining such
an orientation completion (if one exists) is solvable in polynomial time 
\end{corollary}  

\pf By Theorem \ref{friendly-general}, it suffices to check whether 
the input friendly partially oriented graph has a directed cycle in a 
non-universal cell or a directed cycle contained in the in-neighbourhood or 
the out-neighbourhood of a vertex. This can be done in polynomial time. 
When no mentioned directed cycle is contained in the input partially oriented 
graph, the constructive proof of Theorem \ref{friendly-general} explains how 
to complete it to a locally transitive local tournament.
\qed

\begin{theorem}
The problem of extending partial proper circular arc representations of
proper circular arc graphs is solvable in polynomial time.
\end{theorem}

\pf We show how to reduce the problem of extending partial representations of 
proper circular arc graphs to the problem of deciding whether a friendly 
partially oriented graph can be completed to a locally transitive local 
tournament. Since the latter problem is solvable in polynomial time by 
Corollary \ref{friendly-ext} so is the former one. 
Let $G = (V,E)$ be a proper circular arc graph and $H$ be an induced subgraph 
of $G$.
Suppose that $J_v, v \in V(H)$ is a proper circular arc representation of $H$
(i.e., a partial proper circular arc representation of $G$). We obtain an
orientation of $H$ in such a way that $(u,v)$ is an arc if and only if
$J_u$ contains the counterclockwise endpoint of $J_v$. 
Let $C_1, C_2, \dots, C_r$ be the connected components of $\overline{G}$,
each of which contains at least one vertex from $H$.
Let $H'$ be the subgraph of $G$ induced by $V(C_1) \cup V(C_2) \cup \cdots 
\cup V(C_r)$. We extend the orientation from $H$ to $H'$ as follows:
We first orient all unoriented edges in the cells of $H'$ so that each 
becomes transitive tournament. For each $i = 1, 2, \dots, r$, if 
any unbalanced edge within $C_i$ is an oriented edge, then we extend 
the orientation of $H$ to all unbalanced edges of $G$ within $C_i$ 
using a colour class of the corresponding thick component of $G^+$
(see Theorem \ref{implication-class}).
For each pair $i, j$, there must be at least one oriented edge between
$C_i$ and $C_j$, we extend the orientation of $H$ to all edges of $G$ 
between $C_i, C_j$ using a colour class of the corresponding thick component 
of $G^+$ (see Theorem \ref{implication-class}). 
Thus we obtain a partial orientation of $H'$, which yields a partial orientation 
$P = (V,A \cup E')$ of $G$ where $A$ consists of all arcs in $H'$. 
The definition of $H'$ and the orientation of $H'$ ensures that $P$
is consentaneous. It also implies that in any triangle of $G$, 
if the three edges in the triangle correspond to vertices from three different 
connected components in $G^+$, then they are all oriented in $H'$ and hence
in $P$. Therefore $P$ is friendly. It remains to show that the proper 
circular arc representation of $H$ can be extended to a proper circular arc
representation of $G$ if and only if $P$ can be completed to a locally
transitive local tournament. Suppose that $G$ has a proper circular arc
representation which extends the proper circular arc representation of $H$.
Then we can obtain an orientation of $G$ that is a locally transitive local
tournament using the representation of $G$ (in a similar way as above for $H$). 
By Theorem \ref{implication-class} and the defnition of $P$, the orientation 
of $G$ is an orientation completion of $P$. Conversely, suppose that 
$P$ is completed to a locally transitive local tournament. Then 
it is possible to extend the proper circular arc representation of $H$ 
to a proper circular arc representation of $G$ 
(see Theorem~3.1 in \cite{hellJGT20}).
\qed

\section{Remarks and open problems}
\label{remarks}

A digraph $D=(V,A)$ is {\bf $k$-arc-strong} for some $k>0$ if it remains strongly connected after the deletion of any subset $A'\subseteq A$ of at most $k-1$ arcs. A digraph $D=(V,A)$ is {\bf $k$-strong} if $|V|\geq k+1$ and $D-X$ is strong for every subset $X\subset V$ of size at most $k-1$.

As we mentioned in the introduction, the orientation completion problem is polynomially solvable for the class of strong digraphs. It is natural to ask about the complexity of the problem for the class of $k$-arc-strong, respectively the class of $k$-strong digraphs.

\begin{theorem}
  The orientation completion problem is polynomially solvable for the class of $k$-arc-strong digraphs.
\end{theorem}

\pf (Sketch) It is well-known, see e.g \cite[Section 11.8]{bang2009} that there is a polynomial algorithm based on submodular flows for deciding whether a given undirected graph $G$ has a $k$-arc-strong orientation. The way this works is that first an arbitrary orientation $D$ is assigned to the edges of $G$ and then using a submodular flow algorithm, we can determine whether we can reorient some arcs so that the result is $k$-arc-strong (which by Nash-Williams orientation theorem (see eg. \cite[Theorem 11.5.3]{bang2009}) is the case precisely when $G$ is $2k$-edge-connected). Furthermore we can find a set of arcs whose reversal results in a $k$-arc-strong digraph if such a set exists. This approach does not quite work for the orientation completion problem since the starting orientation which we chose has to agree with the directed arcs of the input pog $P$ and none of these may be reversed. The solution is to use a polynomial minimum cost submodular flow algorithm in which the two different orientations of an edge have (possibly different) costs assigned to them and the goal is to find a minimum cost submodular flow, corresponding to a minimum cost set of arcs of $D$ whose reversal leads to a $k$-arc-strong orientation of $G$. Now we just have to assign infinite cost to any arc which is opposite to one of the arcs of $P$ and zero to all other orientations (including both orientations of the edges of $P$), implying that $P$ has a $k$-arc-strong completion if and only if the mincost feasible submodular flow has cost zero. \qed

\begin{theorem} \label{k-conn}
  For any natural number $k\geq 3$ the orientation completion problem is NP-complete for
  the class of $k$-strong digraphs. 
  \end{theorem}

\pf It was shown in \cite{decevignyarXiv1212.4086} that for every natural number $k\geq 3$ it is NP-complete to decide whether an undirected graph has a $k$-striong orientation. Hence also the more general orientation completion problem is NP-complete for the class of $k$-strong digraphs. \qed

Thomassen \cite{thomassenJCT110} proved that a graph $G$ has a 2-strong orientation if and only if $G$ is 4-edge-connected and $G-v$ is 2-edge-connected for every vertex $v$. This implies that the orientation problem for the class of 2-strong digraphs is polynomial. It is also easy to check whether a given digraph is 2-strong. However, as far as we know, the complexity of deciding whether the edges of a mixed graph can be oriented so that the resulting digraph is 2-strong is open. 

\begin{problem}
  What is the complexity of the orientation completion problem for the class of 2-strong digraphs?
  \end{problem}

A digraph is an {\bf in-tournament} if the set of in-neighbours of every vertex induces a tournament.

\begin{theorem}
  \label{completetoInT}
  The orientation completion problem is polynomial for the class of in-tournaments.
  \end{theorem}

\pf In \cite[Section 11.1.4]{bang2009} it is shown how to reduce the problem of deciding whether a graph can be oriented as an in-tournament to an instance of 2-SAT. It is not difficult to see that we may extend that reduction to work when the input is a pog instead of a graph. We leave the details to the interested reader. \qed

\begin{proposition}\cite{bangJCT59}
  \label{chordinT}
  A graph is chordal if and only if it has an orientation as an acyclic in-tournament.
\end{proposition}

\begin{problem}
  What is the complexity of the orientation completion problem for the class of acyclic in-tournaments?
  \end{problem}

We may also ask about the complexity of other properties of the target graph such as having a directed cycle factor, that is, a spanning collection of vertex-disjoint directed cycles. The following result shows that for this class the orientation completion problem is hard.

\begin{theorem}
It is NP-complete to decide whether pog $P$ has a completion $D$ with a directed cycle factor.
\end{theorem}

\pf It was shown in \cite{bangDAM193} that is NP-complete to decide  whether a bipartite digraph $B$ has a directed cycle-factor $C_1,C_2,\ldots{},C_k$ so that no $C_i$ has length 2. Let $B$ be given and form the pog $P$ from $B$ by replacing the two arcs of each directed 2-cycle by an edge. It is easy to see that $P$ has a completion with a directed cycle factor if and only if $B$ has a cycle factor with no directed 2-cycle, implying the theorem. \qed

Let $\pi=\{(s_1,t_1),\ldots{},(s_k,t_k)\}$ be a set of $k$ pairs of  distinct vertices in a (di)graph $H$. A {\bf $\pi$-linkage} in $H$ is a collection of $k$ disjoint paths $R_1,\ldots{},R_k$ so that $R_i$ starts in $s_i$ and ends in $t_i$. For a given class $\cal C$ of digraphs, the 
{\bf ${\cal C}$-$\pi$-linkage completion} problem is as follows: given a pog $P=(V,E\cup A)$ and a set $\pi$ of $k$ terminal pairs in $V$; it is possible to complete the orientation of $P$ so that the resulting oriented graph is in $\cal C$ and has a  $\pi$-linkage?

For general digraphs the $\pi$-linkage problem, and hence also the completion version, is NP-complete already when $k=2$ and even if the digraph is highly connected \cite{fortuneTCS10,thomassenC11}. Chudnovsky, Scott and Seymour \cite{chudnovskyAM270} proved that the $\pi$-linkage problem is polynomial for semicomplete digraphs (that is digraphs whose underlying graph is complete), This implies that the {\bf tournament-$\pi$-linkage completion} problem is polynomial because such a completion is possible if and only if the digraph that we obtain from $P$ by replacing 
each undirected edge by a directed 2-cycle is semicomplete and has a $\pi$-linkage (no two paths in a linkage intersect).

  When  $k$ (the number of pairs to be linked) is part of the input, the {\bf acyclic-$\pi$-linkage completion} problem  is NP-complete already when $P$ is an acyclic oriented graph or just a graph  the problem is NP-complete  \cite{fortuneTCS10,lynchSIGDA5}.

\begin{problem}
\label{toacyclicklink}
What is the complexity of the acyclic-$\pi$-linkage completion problem when $k$ is fixed?
\end{problem}

  By the Robertson-Seymour linkage algorithm \cite{robertsonJCT63}, the acyclic-$\pi$-linkage completion problem is polynomial when $P$ is a graph: first find, using the $O(n^3)$ algorithm for $k$-linkage in graphs, a set of $k$ disjoint paths $R_1,\ldots{},R_k$ linking the terminals. If no such set exists, we can report that the problem has no solution. Otherwise consider the ordering of $V$ which  lists $V$ as $V(R_1),V(R_2),\dots{},V(R_k),X$, where $X$ are the vertices not on any of the paths. Now just orient all edges  from left to right to obtain an acyclic completion which still contain the paths $R_i$, $i\in [k]$. At the other extreme, when $P$ is already an acyclic oriented graph, we may again decide the problem  in polynomial time as shown by Fortune et all \cite{fortuneTCS10}.

The following generalizes round orderings. A cyclic ordering ${\cal O}=v_1,v_2,\ldots{},v_n,v_1$ is {\bf nice} if each there is no  triple $v_k,v_i,v_j$ where $(v_i,v_k),(v_j,v_i)$ are arcs and the vertices occur in the given order according to $\cal O$. It is easy to check that every excellent cyclic ordering is also nice but the converse need not hold.

\begin{problem}
\label{niceexist}
What is the complexity of deciding whether an oriented graph $D$ has a nice cyclic ordering?
\end{problem}

It was shown in \cite{galilTCS5} that when the input is just a set of cyclic triples (not coming from in-and out-neighbours of a digraph) and the goal is to find a cyclic  ordering which agrees with all the triples, the problem is NP-complete.

\begin{problem}
Does every oriented graph with a nice cyclic ordering also have an excellent cyclic ordering?
\end{problem}

If true, then  Theorem \ref{LTTexthard} would imply that Problem \ref{niceexist} is NP-complete.

\end{document}